\documentclass[amsmath,amssymb,prb,twocolumn,showpacs]{revtex4}
\usepackage[dvips]{graphicx}
\usepackage{bm}% bold math
\usepackage{bm}% bold math

\usepackage{color}
\newcommand{\YU}[1]{#1}

\begin{document}
\title{Full-counting statistics of information content in the presence of Coulomb interaction}
\author{Yasuhiro Utsumi}
\address{Department of Physics Engineering, Faculty of Engineering, Mie University, Tsu, Mie, 514-8507, Japan}

\begin{abstract}
We calculate the R\'enyi entropy of a \YU{positive} integer order $M$ for a reduced density matrix of a single-level quantum dot connected to left and right leads. 
We exploit a $2 \times 2$ modified Keldysh Green function matrix obtained by the discrete Fourier transform of a $2 M \times 2M$ multi-contour Keldysh Green function matrix. 
%[Yu. V. Nazarov, Phys. Rev. B {\bf 84}, 205437 (2011)]. 
A moment generating function of self-information is deduced from the analytic continuation of $M$ to the complex plane. 
We calculate the probability distribution of self-information and find that, within the Hartree approximation, the on-site Coulomb interaction affects rare events and modifies a bound of the probability distribution. 
A simple equality, from which an upper bound of the average, i.e., the entanglement entropy, would be inferred, is presented. 
For noninteracting electrons, the entanglement entropy is expressed with current cumulants of the full-counting statistics of electron transport. 
\end{abstract}

\date{\today}


%PACS

%05.40.-a 	Fluctuation phenomena, random processes, noise, and Brownian motion (for fluctuations in superconductivity, see 74.40.-n; for statistical theory and fluctuations in nuclear reactions, see 24.60.-k; for fluctuations in plasma, see 52.25.Gj; for nonlinear dynamics and chaos, see 05.45.-a)

%72.70.+m	Noise processes and phenomena

%73.63.Kv 	Quantum dots

%73.23.-b 	Electronic transport in mesoscopic systems

%05.30.-d 	Quantum statistical mechanics (for quantum fluids aspects, see 67.10.Fj)

%03.67.Mn 	Entanglement measures, witnesses, and other characterizations (see also 03.65.Ud Entanglement and quantum nonlocality; 42.50.Dv Quantum state engineering and measurements in quantum optics)

%03.67.-a 	Quantum information (see also 42.50.Dv Quantum state engineering and measurements; 42.50.Ex Optical implementations of quantum information processing and transfer in quantum optics)
 
%65.40.gd 	Entropy

%05.30.-d 	Quantum statistical mechanics (for quantum fluids aspects, see 67.10.Fj)

%89.70.-a 	Information and communication theory (for telecommunications, see 84.40.Ua; for optical communications, see 42.79.Sz; for quantum information, see 03.67.-a; for applications to neuroscience, see 87.19.lo)
 
%89.70.Cf 	Entropy and other measures of information

%89.70.Kn 	Channel capacity and error-correcting codes
 
%03.67.Hk 	Quantum communication

%05.70.Ln 	Nonequilibrium and irreversible thermodynamics (see also 82.40.Bj Oscillations, chaos, and bifurcations in physical chemistry and chemical physics)

%3.23.Hk 	Coulomb blockade; single-electron tunneling

\maketitle

\newcommand{\mat}[1]{\mbox{\boldmath$#1$}}

\section{Introduction}

Full-counting statistics is a powerful theoretical tool to investigate the statistical properties of electron transport~\cite{LLL,NazarovBook}. 
This statistical method enables us to calculate the probability distribution of the number of electrons transferred between two subsystems, left and right leads connected by a quantum conductor. 
The entanglement entropy is also a measure of correlations between the two subsystems~\cite{Beenakker,KL,FrancisSong1,FrancisSong2,Petrescu,Thomas}. 
Suppose we partition our total system into complementary subsystems $A$ (the left lead and the quantum conductor) and B (the right lead). 
Then the partial trace of a density matrix of the total system $\rho$ over the subsystem $B$ degrees of freedom, 
%------------------------------------------------------------------------------
\begin{align}
\rho_A
=
{\rm Tr}_B
\rho
\, , 
\end{align}
%------------------------------------------------------------------------------
defines the reduced density matrix. 
The operator of the information content, i.e., the self-information associated with an outcome described by the reduced density matrix, may be given by 
%------------------------------------------------------------------------------
$I=-\ln \rho_A$ (we choose base $e$). 
%------------------------------------------------------------------------------
\YU{
The operator is often called the entanglement Hamiltonian and its spectrum, the entanglement spectrum~\cite{Li}, has been widely used to study topological phases. 
} 
The entanglement entropy is the von Neumann entropy~\cite{NC} of the reduced density matrix given as
%------------------------------------------------------------------------------
\begin{align}
\langle I \rangle
=
{\rm Tr}_A
\left[
\rho_A I
\right]
\, , 
\label{entent}
\end{align}
%------------------------------------------------------------------------------
where ${\rm Tr}_A$ means the partial trace over the subsystem $A$ degrees of freedom. 
Technically, it is convenient to exploit the ``R\'enyi entropy" of order $M$~\cite{Cover,note1}, 
%------------------------------------------------------------------------------
\begin{align}
S_M
=
{\rm Tr}_A
\left[
{\rho_A}^M
\right]
\, , 
\label{renyi}
\end{align}
%------------------------------------------------------------------------------
and calculate the entanglement entropy from its derivative
$\langle I \rangle = - \lim_{M \to 1} \partial S_M/\partial M$ 
\YU{
(
Precisely, Eq. (\ref{renyi}) is a modified definition introduced in Ref.~\onlinecite{Nazarov}, 
which is convenient for our purpose
).
}

The entanglement entropy~\cite{KL} and the R\'enyi entropy~\cite{FrancisSong2} are closely related to the Levitov-Lesovik formula~\cite{Klich}, the current cumulant generating function of the full-counting statistics: They are expressed by a unique quantity, the correlation matrix~\cite{AI}. 
However, the relation is limited to non-interacting electrons. 
Recently, Nazarov proposed another approach to calculate the R\'enyi entropy of an %\YU{positive} 
integer order $M$ by introducing the Keldysh Green function defined on a multi-contour, which is a sequence of $M$ replicas of a standard Keldysh contour~\cite{Nazarov}. 
Ansari and Nazarov have further developed this method~\cite{Ansari1,Ansari2,Ansari3} and relate the flow of R\'enyi entropy with the flow of heat~\cite{Ansari2}. 
Although the results are limited to weak coupling between the two subsystems, the approach would be promising since it enables one to utilize field theory techniques.

\begin{figure}[ht]
\includegraphics[width=0.5 \columnwidth]{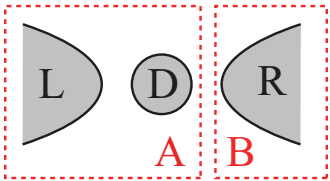}
\caption{
The single-level quantum dot coupled to left and right leads. 
We partition the total system into subsystems $A$ and $B$ and trace out the degrees of freedom associated with the subsystem $B$. 
The subsystem $A$ consists of the dot and the left lead. 
The subsystem $B$ is the right lead. 
}
\label{setup}
\end{figure}

In the present paper, we consider a single-level quantum dot connected to left and right leads [Fig.~\ref{setup}] and calculate the R\'enyi entropy 
\YU{of a positive integer order $M$} 
by accounting for the tunnel coupling to all orders as well as the on-site Coulomb interaction up to the lowest order. 
The reduced density matrix is derived by tracing out the degrees of freedom associated with the right lead. 
We will utilize the anti-periodicity of the multi-contour Keldysh Green function and perform the discrete Fourier transform. 
The `Matsubara frequency'~\cite{ETU,AGD} introduced in this way 
%------------------------------------------------------------------------------
\begin{align}
\lambda_\ell
=
\pi \left(1 - \frac{2\ell+1}{M} \right)
\, ,
\;\;\;\;
(\ell=0,\cdots,M-1)
\, , 
\label{discou}
\end{align}
%------------------------------------------------------------------------------
is a `counting field'~\cite{NazarovBook}, which counts an electron transfer between replicated Keldysh contours. 
The resulting $2 \times 2$ modified Keldysh Green function matrix is closely related to that previously introduced in the context of the full-counting statistics [see, e.g. Refs.~\onlinecite{NazarovBook,UGS,GK,U,SU,US,HNB,UEUA,Esposito} and references therein]. 
This enables us to apply the Keldysh diagrammatic technique to calculate the R\'enyi entropy, which is formally a `Keldysh partition function' defined on the multi-contour. 
For non-interacting electrons, we relate the R\'enyi entropy with the current cumulants, Eq.~(\ref{universal1}), as previously demonstrated by Song {\it et al.}~\cite{FrancisSong2} based on the correlation matrix. 

Another purpose of the present paper is to examine the idea of the full-counting statistics of self-information $I$. 
After the analytic continuation of $M \to 1- i \xi$, the R\'enyi entropy turns into the information generating function~\cite{Golomb,Guiasu}, which is the moment generating function of probability distribution of self-information. 
Although it can generate all orders of moments, the statistical properties of moments  other than the first moment (\ref{entent}) have rarely been investigated. 
We herein calculate the probability distribution of self-information and find that although the on-site Coulomb interaction weakly affects the entanglement entropy, it also affects rare events and alters a bound of the probability distribution. 
We also note a simple equality~(\ref{Jarzynski1}) from which the upper bound of the entanglement entropy~(\ref{Jarzynski2}) can be deduced. 

In the following, we concentrate on the entanglement entropy under the DC source-drain bias voltage in the limit of long measurement time. 
For non-interacting electrons, our result reproduces the expression by Beenakker~\cite{Beenakker}, Eq.~(\ref{uncentent}). 
In order to estimate the accessible entanglement, one has to perform the projection measurement on electron numbers of subsystems, which only generates a sub-leading correction~\cite{Klich1}. 

%----------------------------------------------------------
%----------------------------------------------------------

\YU{
Here, we re-emphasize two messages, which we feel the most important in the present paper; 
(1) We will introduce the discrete Fourier transform of the multi-contour Keldysh Green function defined in Refs.~\onlinecite{Nazarov,Ansari1,Ansari2,Ansari3}. 
It provides a feasible way to calculate the R\'enyi entropy starting from a microscopic model Hamiltonian. 
It enables us to treat electron interactions with a slight extension of traditional diagrammatic techniques. 
(2) Through explicit calculations, we would like to demonstrate that the R\'enyi entropy may contain the information on the fluctuations beyond the entanglement entropy. 
We expect that the concept of the probability distribution of self-information could provide a way to interpret the meaning of the R\'enyi entropy. 
}

The paper is organized as follows. 
We summarize the information generating function in Sec.~\ref{igf} and then introduce our model Hamiltonian in Sec.~\ref{model}. 
In Sec.~\ref{sec:Ren}, after we analyze the R\'enyi entropy for decoupled systems, we express the R\'enyi entropy in the form of the `Keldysh partition function' defined on the multi-contour. 
Then, we summarize the discrete Fourier transform of the modified Keldysh Green function. 
In Sec.~\ref{sec:nonele}, we present the results for noninteracting electrons. 
In Sec.~\ref{CI}, we discuss the effect of Coulomb interaction within the Hartree approximation. 
Section~\ref{Sum} summarizes our findings.

\section{Information generating function}
\label{igf}

The information generating function~\cite{Golomb,Guiasu}, the moment generating function of the self-information $I$, would be obtained from the R\'enyi entropy~(\ref{renyi}) by extending 
%a \YU{positive} integer 
$M$ to a complex value $1-i \xi$: 
%------------------------------------------------------------------------------
\begin{align}
S_{1- i \xi}
=
\left \langle 
\rho_A(\tau)^{-i \xi}
\right \rangle
\equiv
\int dI
e^{i \xi I}
P_\tau(I)
\, . 
\label{infgenfun}
\end{align}
%------------------------------------------------------------------------------
\YU{
Here, $\tau$ is a measurement time. 
}
The information generating function satisfies the normalization condition 
$S_{1}=\int dI P_\tau(I)=1$. 
The $n$-th cumulant is 
%------------------------------------------------------------------------------
\begin{align}
\langle \! \langle I^n \rangle \! \rangle
=
\left. 
\frac{\partial^n
\ln S_{1-i\xi}
}{\partial (i \xi)^n}
\right|_{\xi=0}
\, . 
\label{nthcumulant}
\end{align}
%------------------------------------------------------------------------------
The first cumulant is the entanglement entropy (\ref{entent}), 
$
\langle I \rangle = \langle \! \langle I \rangle \! \rangle
$. 
The second cumulant (variance) is 
%------------------------------------------------------------------------------
$
\langle \! \langle I^2 \rangle \! \rangle
=
\langle I^2 \rangle
-
{\langle I \rangle}^2
$. 
%------------------------------------------------------------------------------
The probability distribution of the self-information may be obtained by the inverse Fourier transform, 
%------------------------------------------------------------------------------
\begin{align}
P_\tau(I)
=
\int \frac{d \xi}{2 \pi}
S_{1-i \xi}
e^{-i \xi I}
=
\langle 
\delta(I+\ln \rho_A(\tau))
\rangle
\, . 
\label{iftpi}
\end{align}
%------------------------------------------------------------------------------
The fluctuations are induced by degrees of freedom associated with the subsystem $B$, which have been traced out. 
We note a simple identity, which is reminiscent of the Jarzynski equality~\cite{Esposito,Jarzynski,Campisi}, 
\YU{
%------------------------------------------------------------------------------
\begin{align}
\left \langle e^{I} \right \rangle
=
\int dI P(I)e^{I}
=
S_0
=
{\rm Tr}_A \left[ {\rho_A(\tau)}^0 \right]
\, , 
\label{Jarzynski1}
\end{align}
%------------------------------------------------------------------------------
which would follow  from Eq.~(\ref{infgenfun}) by setting $\xi=-i$. 
The RHS of Eq. (\ref{Jarzynski1}) is the rank of ${\rho_A(\tau)}^0$, 
which is the unit matrix in the {\it available} many-body Fock space 
[see Ref.~\onlinecite{noteres1}]. 
}
If $P (I) \geq 0$, Jensen's inequality provides the upper bound of the entanglement entropy, 
\YU{
%------------------------------------------------------------------------------
\begin{align}
\ln {\rm Tr}_A 
\left[
{\rho_A(\tau)}^0
\right]
\geq
\left \langle {I} \right \rangle
\, , 
\label{Jarzynski2}
\end{align}
%------------------------------------------------------------------------------
}
which means that the entanglement entropy cannot exceed the entropy of the uniform distribution over available states in the many-body Fock space of subsystem $A$.

\section{Model}
\label{model}

The Hamiltonian of the single-level quantum dot connected to left and right leads is 
%------------------------------------------------------------------------------
\begin{align}
H = 
H_L + H_R + H_D
+
H_T
+
H_{\rm int}
\, , 
\label{hamiltonian}
\end{align}
%------------------------------------------------------------------------------
The quantum dot is represented by a localized level with the energy $\epsilon_D$, 
%------------------------------------------------------------------------------
\begin{align}
H_D = \sum_{\sigma} \epsilon_D d_\sigma^\dagger d_\sigma
\, , 
\end{align}
%------------------------------------------------------------------------------
where $d_\sigma$ is an annihilation operator of an electron with spin $\sigma$. 
The on-site Coulomb interaction is given by, 
%------------------------------------------------------------------------------
\begin{align}
H_{\rm int} &= 
U 
d_{\downarrow}^\dagger
d_{\uparrow}^\dagger
d_{\uparrow}
d_{\downarrow}
\, , 
\end{align}
%------------------------------------------------------------------------------
The left ($r=L$) and right ($r=R$) leads are described by the free electron gas as 
%------------------------------------------------------------------------------
\begin{align}
H_r = 
\sum_{k \sigma} \epsilon_{rk} a_{r k \sigma}^\dagger a_{r k \sigma} 
\, , 
\end{align}
%------------------------------------------------------------------------------
where $a_{r k \sigma}$ annihilates an electron with wave number $k$ and spin $\sigma$. 
The tunneling between the dot and the lead $r$ is described by 
%------------------------------------------------------------------------------
\begin{align}
H_T = 
\sum_{r=L,R} 
\sum_{k \sigma} 
J_r d_\sigma^\dagger a_{r k \sigma} + {\rm H.c.} 
\end{align}
%------------------------------------------------------------------------------
The tunnel coupling broadens the DOS of the quantum dot: 
%------------------------------------------------------------------------------
\begin{align}
\rho(\omega)
=
\frac{\Gamma}{2 \pi [(\omega - \epsilon_D)^2 + \Gamma^2/4]}
\, ,
\;\;\;\;
\Gamma
= 
\Gamma_L
+
\Gamma_R
\, . 
\end{align}
%------------------------------------------------------------------------------
The coupling strength 
$\Gamma_r = 2 \pi \sum_k |J_r|^2 \delta (\omega - \epsilon_{rk})$
between the quantum dot and the lead $r$ is assumed to be energy independent. 
The transmission probability through the quantum dot is proportional to the DOS as 
%------------------------------------------------------------------------------
\begin{align}
{\mathcal T}(\omega)
=
2 \pi 
\frac{\Gamma_L \Gamma_R}{\Gamma}
\rho(\omega)
\, . 
\end{align}
%------------------------------------------------------------------------------

We assume that initially the dot and the leads are decoupled and the Coulomb interaction is switched off. 
Then electrons in each region are equilibrated with the inverse temperature $\beta=1/(k_{\rm B} T)$ (We set $\hbar=k_{\rm B}=e=1$). 
The initial equilibrium density matrix is 
%------------------------------------------------------------------------------
$
\rho_{\rm eq}
=
\rho_{L}
\rho_{R}
\rho_{D}
$, 
where
%------------------------------------------------------------------------------
\begin{align}
\rho_{D}
=&
e^{-\beta (H_D - \mu_D \sum_{\sigma} d_{\sigma}^\dagger d_{\sigma})}
/Z_{D}
\, , 
\\
\rho_{r}
=&
e^{-\beta (H_r - \mu_r \sum_{k \sigma} a_{r k \sigma}^\dagger a_{r k \sigma} )}
/Z_{r}
\, , 
\;\;
(r=L,R) \, , 
\end{align}
%------------------------------------------------------------------------------
are equilibrium density matrices of the quantum dot and lead $r$.
Here $\mu_r$ ($r=L,R,D$) is the chemical potential. 
The equilibrium partition function $Z_{r}$ ensures ${\rm Tr} \rho_{r}=1$. 
In each region, an electron and a hole obey the following distribution functions: 
%------------------------------------------------------------------------------
\begin{align}
f_{r}^+(\omega)
=
\frac{1}
{1+e^{\beta (\omega-\mu_r)} }
\, ,
\;\;\;\;
f_r^-(\omega) = 1 - f_r^+(\omega)
\, . 
\end{align}
%------------------------------------------------------------------------------

\section{R\'enyi entropy}
\label{sec:Ren}

\subsection{Two zero temperature limits}
\label{zerlim}

For a warm-up, let us calculate the R\'enyi entropy of a \YU{positive} integer order $M$ for the initial equilibrium density matrix, 
%------------------------------------------------------------------------------
$
s_M 
=
{\rm Tr}_{LD}
\left[ 
{{\rm Tr}_R {\rho_{{\rm eq}}}}^M
\right]
=
{\rm Tr}_D  \left[ {\rho_{D}}^M \right]
{\rm Tr}_L \left[ {\rho_{L}}^M \right]
=
\prod_\sigma
s_{D \sigma M}
s_{L \sigma M}
$. 
%------------------------------------------------------------------------------
Spin-resolved R\'enyi entropies of the quantum dot and left lead are 
%------------------------------------------------------------------------------
\begin{align}
s_{D \sigma M}
=&
\sum_\pm
f_D^\pm(\epsilon_D)^M
\, , 
\label{renyqd}
\\
s_{L \sigma M}
=&
\prod_{k}
\sum_\pm 
f_L^\pm(\epsilon_{Lk})^M
\, . 
\end{align}
%------------------------------------------------------------------------------
Let us focus on the quantum dot (\ref{renyqd}) and evaluate the probability distribution by the analytic continuation $M \to 1-i \xi$ and then the inverse Fourier transform (\ref{iftpi}). 
At a finite temperature, we obtain 
%------------------------------------------------------------------------------
\begin{align}
P_{D \sigma}(I)
=
\sum_\pm f_{D}^\pm(\epsilon_D) 
\, \delta(I+\ln f_{D}^\pm(\epsilon_D)) \, , 
\end{align}
%------------------------------------------------------------------------------
which satisfies 
$\langle e^I \rangle = \int dI P_{D \sigma}(I) e^I =2$
for an arbitrary temperature. 
%On the other hand, at zero temperature, since $s_{D \sigma M}=1$, we obtain $P_{D \sigma}(I) = \delta(I)$, which results in $\langle e^I \rangle =1$. The above discussions suggest that when we apply the equality (\ref{Jarzynski1}) to the zero temperature case, we have to specify the procedures  {\bf (I)} and {\bf (II)}. 
\YU{
Although the equality would be valid at zero temperature, we have to pay attention to zero temperature limits and specify the procedures {\bf (I)} and {\bf (II)}. 
}

{\bf (I)} 
We first take a zero temperature limit for a \YU{positive} integer $M=1,2,\cdots $
and then perform the analytic continuation $M \to 1-i \xi$. 
For the initial equilibrium density matrix, the resulting information generating function is $s_{1-i \xi}=1$ since 
\YU{$\sum_\pm [f^\pm_r(\omega)]^M = 0^M+1^M=1$ ($r=L,D$)
at zero temperature for a positive integer $M$.
} 
The probability distribution is then $P_\tau(I)=\delta(I)$ and the equality (\ref{Jarzynski1}) is $\langle e^I \rangle =1$. 

{\bf (II)}
We first perform the analytic continuation at a finite temperature. 
For the initial equilibrium density matrix, since $[{f_r^\pm}(\omega)]^0=1$ 
\YU{at a finite temperature}, the equality ~(\ref{Jarzynski1}) leads to $\langle e^I \rangle = 2^{N_A}$, where $N_A={\sum_\sigma 1+\sum_{k \sigma}1}$. 
This provides the maximum entropy of the subsystem $A$, $N_A \ln 2$. 
The average is the thermodynamic entropy, 
%------------------------------------------------------------------------------
\begin{align}
\langle I \rangle
=&
S_A
=
-
\sum_\sigma
\sum_\pm f_D^\pm (\epsilon_D) \ln f_D^\pm (\epsilon_D)
\nonumber \\
&-
\sum_{k \sigma}
\sum_\pm f_L^\pm (\epsilon_{Lk}) \ln f_L^\pm (\epsilon_{Lk})
\, , 
\end{align}
%------------------------------------------------------------------------------
which vanishes at zero temperature. 
The inequality~(\ref{Jarzynski2}) ensures that the relative entropy~\cite{Cover} between the equilibrium distribution and the uniform distribution is non-negative.

\subsection{Replica method}
\label{rtre}

Following Refs. \onlinecite{Nazarov,Ansari1,Ansari2,Ansari3}, we formulate the perturbation theory of the R\'enyi entropy~(\ref{renyi}) of the full-density matrix at time $\tau$: 
%------------------------------------------------------------------------------
\begin{align}
\rho(\tau)
= 
U(\tau)  \rho_{\rm eq} U(\tau)^\dagger
\, ,
\;\;\;\;
U(\tau) = e^{-i H \tau}
\, . 
\label{rhofull}
\end{align}
%------------------------------------------------------------------------------
We treat the tunnel Hamiltonian and the on-site Coulomb interaction as the perturbation, 
%------------------------------------------------------------------------------
$
V = H_T + H_{\rm int} 
$, 
%------------------------------------------------------------------------------
and rewrite the Hamiltonian (\ref{hamiltonian}) as $H=H_0+V$. 
The R\'enyi entropy~(\ref{renyi}) includes the partial trace over the subsystem $B$, ${\rm Tr}_B$, inside the partial trace over the subsystem $A$, ${\rm Tr}_A$. 
To avoid this complication, we adopt the replica method~\cite{Sherrington}. 
We introduce $M$ ($M$ is a \YU{positive} integer) replicas of subsystem $B$ electron annihilation and creation operators: 
%------------------------------------------------------------------------------
\begin{align}
a_{Rk\sigma} 
\to 
a_{Rk\sigma m}
\, ,
\;\;\;\;
{a_{Rk\sigma}}^\dagger
\to 
{a_{Rk\sigma m}}^\dagger
\, ,
\label{replica}
\end{align}
%------------------------------------------------------------------------------
where $m=1, \cdots, M$. 
Then the Hamiltonian $H$ and the density matrix $\rho_{{\rm eq}}$ are also replicated by the replacement (\ref{replica}). 
We introduce another subscript $m$ to specify $m$-th replicated operators, i.e., $H_{0 \, m}$, $H_{m}$, $U_m$, $V_m$ and $\rho_{{\rm eq} \, m}$. 
The R\'enyi entropy (\ref{renyi}) is expressed by a trace over the total system, the subsystem $A$ plus $M$-replicas of subsystem $B$ as 
%------------------------------------------------------------------------------
\begin{align}
S_M
=&
{\rm Tr}
\left[
U_M
\rho_{{\rm eq} \, M}
U_M^\dagger
U_{M-1}
\rho_{{\rm eq} \, {M-1}}
U_{M-1}^\dagger
\right. 
\nonumber \\
& \times
\left.
\cdots
\times 
U_{2}
\rho_{{\rm eq} \, 2}
U_{2}^\dagger
U_{1}
\rho_{{\rm eq} \, 1}
U_{1}^\dagger
\right]
\nonumber \\
=&
{\rm Tr}
\left[
{U_M}_I
\rho_{{\rm eq} \, M}
{U_M}_I^\dagger
{U_{M-1}}_I
\rho_{{\rm eq} \, M-1}
{U_{M-1}}_I^\dagger
\right. 
\nonumber \\
& \times
\left.
\cdots
\times
{U_2}_I
\rho_{{\rm eq} \, 2}
{U_2}_I^\dagger
{U_1}_I
\rho_{{\rm eq} \, 1}
{U_1}_I^\dagger
\right]
\, , 
\end{align}
%------------------------------------------------------------------------------
where the subscript $I$ indicates the interaction picture 
\YU{$U_{m \, I}=e^{i H_{0 \, m} \tau} {U_m}$. }

\begin{figure}[hb]
\includegraphics[width=0.8 \columnwidth]{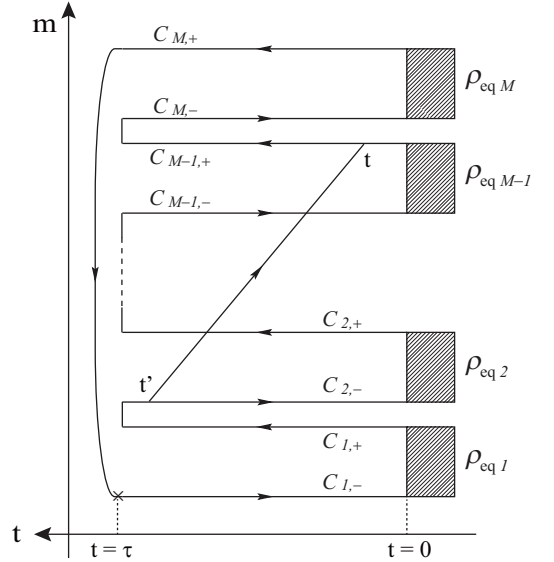}
\caption{
Multi-contour $C$, which represents successive connections of the $M$ replicated Keldysh contours, starting at $t=\tau$ on the lower branch of the first replica $C_{1 , -}$. 
 Shaded boxes are $M$ replicas of the initial equilibrium density matrix $\rho_{{\rm eq} \, m}$ ($m=1,\cdots,M$). 
The solid line connecting $C_{2 , -}$ and $C_{M-1 , +}$ indicates a `greater' Green function $g_{Lk \sigma}^{M-1 \, +,2 \, -}(t,t')$. 
}
\label{rkc}
\end{figure}

The time evolution operator and its Hermite conjugate are expanded as 
%------------------------------------------------------------------------------
$
U_{m \, I}
=
T 
\exp
\left(
-i \int_0^\tau dt 
V_m(t)_I
\right)
$
%------------------------------------------------------------------------------
and 
%------------------------------------------------------------------------------
$
U_{m \, I}^\dagger
=
\tilde{T} 
\exp
\left(
i \int_0^\tau dt 
V_m(t)_I
\right)
\, , 
$
%------------------------------------------------------------------------------
where $T$ and $\tilde{T}$ are the time-ordering and anti-time-ordering operators. 
Here the perturbation Hamiltonian in the interaction picture is 
$
V_m(t)_I
=
e^{i H_{0 \, m} t}
V_m
e^{-i H_{0 \, m} t}
$. 
Then, following Ref.~\onlinecite{Nazarov}, we introduce the multi-contour $C$, which is a sequence of $M$ replicas of the standard Keldysh contour as depicted in Fig.~\ref{rkc}. 
We set a starting point at $t=\tau$ on the lower branch of the first replica $C_{1 , -}$. 
The contour goes to $\rho_{{\rm eq} \, 1}$ at $t=0$ along $C_{1 , -}$ and returns to $t=\tau$ along $C_{1 , +}$. 
Then it connects to $t=\tau$ on the lower branch of the second replica $C_{2 , -}$. 
It successively repeats until it reaches $t=\tau$ on $C_{M , +}$. 
Then it connects to the starting point $t=\tau$ on $C_{1 , -}$. 
By introducing the contour ordered operator $T_C$, the R\'enyi entropy can be expressed as the `Keldysh partition function', 
%------------------------------------------------------------------------------
\begin{align}
S_M
=&
{\rm Tr}
\left[
T_C
\exp
\left(
-i \int_C dt 
V(t)_I
\right)
\rho_{{\rm eq} \, M}
\cdots
\rho_{{\rm eq} \, 1}
\right]
\\
=&
\left \langle 
T_C
\exp
\left(
-i \int_C dt 
V(t)_I
\right)
\right \rangle_M s_M
\, ,
\label{kpf}
\end{align}
%------------------------------------------------------------------------------
where the integral over $t$ is performed along the multi-contour $C$. 
In the following, we sometimes write the time $t$ defined on $C_{m , s}$ as $t_{ms}$ ($s=\pm$). 
Then the explicit form of the operator $V(t)_I$ in Eq.~(\ref{kpf}) is $V(t_{ms})_I = V_{m} (t)_I$. 
Note that the contour-ordering operator $T_C$ also acts on  $\rho_{{\rm eq} \, m} $ residing at $t=0_ {m \pm}$. 
The normalized expectation value is defined as 
%------------------------------------------------------------------------------
\begin{align}
\langle {\mathcal O} \rangle_M
=
{\rm Tr}
\left[
{\mathcal O} 
\rho_{{\rm eq} \, M}
\cdots
\rho_{{\rm eq} \, 1}
\right]
/s_M
\, , 
\label{nev}
\end{align}
%------------------------------------------------------------------------------
where the denominator of the RHS is the R\'enyi entropy of the initial equilibrium density matrix. 
This enables us to exploit the Bloch-De Dominicis theorem~\cite{KTH} (Appendix \ref{KGF}) and the linked cluster theorem, which result in 
%------------------------------------------------------------------------------
\begin{align}
\ln 
\frac{S_M}{s_M}
=&
\sum_{n=1}^\infty
\frac{(-i)^n}{n!}
\int_C 
d t_n \cdots d t_1
\left \langle 
T_C
V(t_n)_I
\cdots 
\right.
\nonumber \\
& \times
\left.
V(t_1)_I
\right \rangle_{M,{\rm c}}
\, . 
\label{lce}
\end{align}
%------------------------------------------------------------------------------
The subscript $c$ means that only connected diagrams are taken into account. 
Equation (\ref{lce}) is the starting point of the following calculations.

%----------------------------------------------------------
%----------------------------------------------------------

\subsection{Modified Keldysh Green function}
\label{mkgf}

The diagrammatic expansion of the Keldysh partition function (\ref{kpf}) is performed based on the multi-contour Keldysh Green function~\cite{Nazarov,Ansari1,Ansari2,Ansari3}. 
We relegate the details to Appendix  \ref{sec:dft} and summarize the multi-contour Keldysh Green function for an electron in the left lead, which is a part of the subsystem $A$. 
This is a correlation function of $a_{Lk\sigma}^\dagger$ on $C_{m' , s'}$ and $a_{Lk\sigma}$ on $C_{m , s}$:
%------------------------------------------------------------------------------
\begin{align}
g_{L k \sigma}(t_{ms},t'_{m's'})
=&
g_{L k \sigma}^{ms,m's'}(t,t')
\nonumber \\
=&
-i 
\left \langle 
T_C
a_{L k \sigma}(t_{ms})_I
a_{L k \sigma}^\dagger(t_{m's'}')_I
\right \rangle_M
\, . 
\label{gf}
\end{align}
%------------------------------------------------------------------------------
This is a component  of a $2M \times 2M$ Keldysh Green function matrix ${\mathbf g}_{L k \sigma}(t,t')$ 
[See the explicit form Eq.~(\ref{repkelgf}) in Appendix~\ref{sec:dft}]. 
By exploiting anti-periodicity in the replicated Keldysh space, we perform a discrete Fourier transform~\cite{ETU,AGD}, 
%------------------------------------------------------------------------------
\begin{align}
{\mathbf g}_{L k \sigma}^{\lambda_\ell}(t,t')
=&
\sum_{m-m'=0}^{M-1}
\left[
{\mathbf g}_{L k \sigma}(t,t')
\right]_{m,m'}
e^{i \pi \frac{2 \ell+1}{M} (m-m')}
\label{dft}
\, . 
\end{align}
%------------------------------------------------------------------------------
This is  $2 \times 2$ Green function matrix defined in a single Keldysh space as
%------------------------------------------------------------------------------
\begin{align}
{\mathbf g}_{L k \sigma}^\lambda(t,t')
=&
\left[
\begin{array}{cc}
g_{L k \sigma}^{\lambda, ++}(t,t') & g_{L k \sigma}^{\lambda, +-}(t,t') \\
g_{L k \sigma}^{\lambda, -+}(t,t') & g_{L k \sigma}^{\lambda, --}(t,t')
\end{array}
\right]
\, . 
\label{mkgfl}
\end{align}
%------------------------------------------------------------------------------
The inverse discrete Fourier transform is 
%------------------------------------------------------------------------------
\begin{align}
\left[
{\mathbf g}_{L k \sigma}(t,t')
\right]_{m,m'}
=&
\frac{1}{M}
\sum_{\ell=0}^{M-1}
{\mathbf g}_{L k \sigma}^{\lambda_\ell}(t,t')
e^{-i \pi \frac{2 \ell+1}{M} (m-m')}
\, . 
\label{idft}
\end{align}
%------------------------------------------------------------------------------
The parameter $\lambda_\ell$ (\ref{discou}), the `Matsubara frequency'~\cite{ETU,AGD}, satisfies $\lambda_{M-1-\ell}=-\lambda_\ell$. 
This parameter is the counting field for electron transfer between different replicated Keldysh contours. 
The explicit form of Eq.~(\ref{mkgfl}) is 
%------------------------------------------------------------------------------
\begin{widetext}
\begin{align}
{\mathbf g}_{L k \sigma}^\lambda(t,t')
=
-i e^{-i \epsilon_{Lk}(t-t')}
\left[
\begin{array}{cc}
f_{L,\lambda}^-(\epsilon_{Lk})
\theta (t-t')
-
f_{L,\lambda}^+(\epsilon_{Lk})
\theta (t'-t)
&
f_{L,\lambda}^+(\epsilon_{Lk}) e^{i \lambda} \\
-
f_{L,\lambda}^-(\epsilon_{Lk}) e^{-i \lambda}
&
f_{L,\lambda}^-(\epsilon_{Lk})
\theta (t'-t)
-
f_{L,\lambda}^+(\epsilon_{Lk})
\theta (t-t')
\end{array}
\right]
\, , 
\label{mmkgf}
\end{align}
\end{widetext}
%------------------------------------------------------------------------------
where the modified electron and hole distribution functions are 
%------------------------------------------------------------------------------
\begin{align}
f_{L,\lambda}^+(\epsilon)
=&
\frac{1}
{1+e^{\beta (\epsilon-\mu_L) + i \lambda} }
\, ,
\;\;
f_{L,\lambda}^-(\epsilon)
=
1
-
f_{L,\lambda}^+(\epsilon)
\, . 
\end{align}
%------------------------------------------------------------------------------
In the limit of zero temperature, they are independent of the counting field $\lambda$, 
%------------------------------------------------------------------------------
$
\lim_{\beta \to \infty}
f_{L,\lambda}^\pm(\epsilon)
=
\theta( \pm (\mu_L-\epsilon) )
$. 
%------------------------------------------------------------------------------
Then Eq.~(\ref{mmkgf}) becomes the `modified Keldysh Green function' introduced in the theory of full-counting statistics~\cite{NazarovBook,UGS,UEUA,GK,U,SU,US,Esposito,HNB}. 
This fact enables us to relate the R\'enyi entropy with the full-counting statistics in a novel manner, which does not rely on the correlation matrix ~\cite{KL,FrancisSong1,FrancisSong2,Petrescu,Thomas}. 
The bare modified Keldysh Green function of the quantum dot is given in the same way [see Eq.~(\ref{mmkgfd}) in Appendix \ref{fuldotmodkelgrefun}].

\begin{figure}[ht]
\includegraphics[width=0.8 \columnwidth]{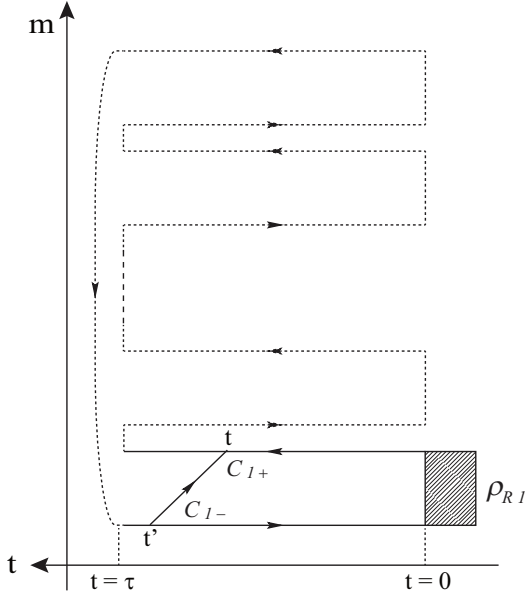}
\caption{
The solid line indicates a `greater' Green function of a replicated fermion $g_{R k \sigma}^{m s,m's'}(t,t')$.
Since it connects the same replica ($m=m'=1$), by deforming the closed time-path, it is reduced to the standard Keldysh Green function except for the choice of a different starting point for the Keldysh contour. 
}
\label{rkcn}
\end{figure}

For an electron in the subsystem $B$, the right lead, replicated annihilation and creation operators, $a_{R k \sigma m}$ and $a_{R k \sigma m}^\dagger$, reside only on the same $m$-th Keldysh contour $C_{m,\pm}$ [see Fig. \ref{rkcn}. The contour $C_{m,\pm}$ starts at $t=\tau_{m,-}$, goes to $t=0_{m,-}=0_{m,+}$ along $C_{m,-}$ and returns to $t=\tau_{m,+}$ along $C_{m,+}$]. 
The multi-contour Keldysh Green function is 
%------------------------------------------------------------------------------
\begin{widetext}
\begin{align}
g_{R k \sigma}^{ms,m's'}(t,t')
=
-i 
\left \langle 
T_C
a_{R k \sigma}(t_{ms})_I
a_{R k \sigma}^\dagger(t_{m s'}')_I
\right \rangle_M
=
-i 
{\rm Tr}
\left[
T_{C_m}
a_{R k \sigma m}(t_{m s})_I
a_{R k \sigma m}^\dagger(t_{m s'}')_I
\rho_{R,m}
\right]
\delta_{m,m'}
\, . 
\label{nrgf}
\end{align}
%------------------------------------------------------------------------------
Here, $T_{C_m}$ is the time-ordering operator along the contour $C_m$. 
The $2 \times 2$ sub-matrix of the $2M \times 2M$ Keldysh Green function matrix is $\left[ {\mathbf g}_{R k \sigma}(t,t') \right]_{m,m'} = {\mathbf g}_{R k \sigma}(t,t') \delta_{m,m'}$, where
%------------------------------------------------------------------------------
\begin{align}
{\mathbf g}_{R k \sigma}(t,t')
=
-i e^{-i \epsilon_{Rk} (t-t')}
\left[
\begin{array}{cc}
f_R^-(\epsilon_{Rk}) \theta(t-t') - f_R^+(\epsilon_{Rk}) \theta(t'-t) & 
f_R^+(\epsilon_{Rk}) \\ 
-f_R^-(\epsilon_{Rk}) & 
f_R^-(\epsilon_{Rk}) \theta(t'-t) - f_R^+(\epsilon_{Rk}) \theta(t-t')
\end{array}
\right]
\, . 
\end{align}
\end{widetext}
%------------------------------------------------------------------------------
This is the same as the standard Keldysh Green function except for the minus signs at off-diagonal components, which are attributed to a different choice of starting point.~\cite{Kamenevbook} 

%----------------------------------------------------------
%----------------------------------------------------------

\section{Noninteracting electrons}
\label{sec:nonele}

\subsection{Linked cluster expansion}
\label{sec:lce}

The linked cluster expansion Eq.~(\ref{lce}) for the non-interacting case $U=0$ can be done straightforwardly~\cite{UGS,U,YIHE}: 
%------------------------------------------------------------------------------
\begin{align}
\ln 
\frac{S_M}{s_M}
&=
\sum_\sigma
\ln 
\frac{S_{\sigma M}}{s_{\sigma M}}
\label{spisum}
\, ,
\\
\ln 
\frac{S_{\sigma M}}{s_{\sigma M}}
&=
-
{\rm Tr}
\left[
g_{D \sigma} \Sigma_\sigma
\right]
-
\frac{1}{2}
{\rm Tr}
\left[
g_{D \sigma} \Sigma_\sigma g_{D \sigma} \Sigma_\sigma
\right]
- \cdots
\label{lincluexp}
\, . 
\end{align}
%------------------------------------------------------------------------------
The self-energy appears after tracing out the degrees of freedom associated with the left and right leads:
%------------------------------------------------------------------------------
$
\Sigma_\sigma(t,t')
=
\sum_{rk}
|J_r|^2
g_{rk \sigma}(t,t') 
\, . 
$
%------------------------------------------------------------------------------
The product and the trace in Eq.~(\ref{lincluexp}) should be understood as the integral along the multi-contour $C$, e.g.,
%------------------------------------------------------------------------------
\begin{align}
{\rm Tr}
\left[
g_{D \sigma} \Sigma_\sigma
\right]
=
\int_C
d t_2 d t_1
g_{D \sigma}(t_1,t_2) \Sigma_\sigma(t_2,t_1) 
\, . 
\end{align}
%------------------------------------------------------------------------------
After we project the time defined on $C$ onto the real axis and perform the discrete Fourier transform, we obtain
%------------------------------------------------------------------------------
\begin{align}
\ln 
\frac{S_{\sigma M}}{s_{\sigma M}}
=&
\sum_{\ell =0}^{M-1}
{\mathcal W}_{\sigma \tau}(\lambda_\ell)
\, , 
\label{maires}
\\
{\mathcal W}_{\sigma \tau}(\lambda)
=&
{\rm Tr}
\ln \left[
{\mathbf 1}-
{\mathbf g}_{D \sigma}^\lambda
\tau_3
{\mathbf \Sigma}_\sigma^\lambda
\tau_3
\right]
\, , 
\label{1pcgf}
\end{align}
%------------------------------------------------------------------------------
where ${\mathbf 1}={\rm diag}(1,1)$ is a unit matrix and $\tau_3={\rm diag}(1,-1)$ is a Pauli matrix in the $2 \times 2$ Keldysh space. 
The trace is performed over the real time $t \in [0,\tau]$ and the $2 \times 2$ Keldysh space. 
The modified self-energy is 
%------------------------------------------------------------------------------
\begin{align}
{\mathbf \Sigma}_\sigma^\lambda
=
|J_R|^2
\sum_k 
{\mathbf g}_{Rk \sigma}
+
|J_L|^2
\sum_k 
{\mathbf g}_{Lk \sigma}^{\lambda}
\, . 
\label{modsel}
\end{align}
%------------------------------------------------------------------------------
At zero temperature, Eq.~(\ref{1pcgf}) becomes the current cumulant generating function of non-interacting electrons (see, e.g. Refs.~\onlinecite{UGS,U,SU,Esposito} and references therein). 
Therefore, Eq.~(\ref{maires}) connects the R\'enyi entropy and the full-counting statistics. 

In the remainder of this section, we consider the zero temperature limit {\bf (I)} in Sec.~\ref{zerlim}. 
The current cumulant generating function is expanded in powers of $i \lambda$ as 
%------------------------------------------------------------------------------
\begin{align}
\sum_\sigma 
{\mathcal W}_{\sigma \tau}(\lambda)
=
\sum_{n=1}^\infty
\frac{
{\mathcal C}_{\tau,n}
(i \lambda)^n
}{n !}
\, , 
\label{defcum}
\end{align}
%------------------------------------------------------------------------------
where ${\mathcal C}_{\tau,n}$ is a $n$-th current cumulant. 
By plugging Eq.~(\ref{defcum}) into Eqs.~(\ref{spisum}) and (\ref{maires}) and by using the relation 
$\sum_{\ell=0}^{M-1} \lambda_\ell^n=0$ for odd $n$, we obtain 
%------------------------------------------------------------------------------
\begin{align}
\ln {S_M}
=
\sum_{n=1}^\infty
\frac{ {\mathcal C}_{\tau,2n} }{(2 n)!}
\left(
\frac{2 \pi i}{M}
\right)^{2n}
\sum_{\ell=0}^{M-1}
\left(
\ell - \frac{M-1}{2}
\right)^{2n}
. 
\label{universal1}
\end{align}
%------------------------------------------------------------------------------
This equation is consistent with the results of Song {\it et al.}
[Eqs. (2.24) and (2.25) in the supplemental material of Ref.~\onlinecite{FrancisSong2}]. 
Further calculations lead to 
%------------------------------------------------------------------------------
\begin{align}
\ln {S_{M}} =& \sum_{n=1}^\infty \frac{{\mathcal C}_{\tau,2n}}{(2n)!} (-1)^n \left( \frac{2 \pi}{M} \right)^{2 n} (\zeta(-2n,(1-M)/2) \nonumber \\ &- \zeta(-2n,(1+M)/2)) \, , \end{align}
%------------------------------------------------------------------------------
where $\zeta(s,a)=\sum_{\ell=0}^\infty (a+\ell)^{-s}$ is the Hurwitz zeta function. 
The entanglement entropy obtained from this equation formally reproduces the results of Klich and Levitov~\cite{KL}, as demonstrated in Ref.~\onlinecite{FrancisSong2}, 
%------------------------------------------------------------------------------
\begin{align}
\langle I \rangle
=
\sum_{n=1}^\infty
\frac{{\mathcal C}_{\tau,2n}}{(2n)!}
(-1)^{n+1} (2 \pi)^{2 n}
B_{2n}
\, ,
\label{universal2}
\end{align}
%------------------------------------------------------------------------------
where $B_n$ is the Bernoulli number.

Let us consider the Gaussian approximation, i.e., we keep only the lowest cumulant, the second cumulant ${\mathcal C}_{\tau,2}$. 
%------------------------------------------------------------------------------
\begin{align}
\ln {S_{M}}
=
\frac{ {\mathcal C}_{\tau,2} }{2}
\sum_{\ell=0}^{M-1} (i \lambda_\ell)^2
=
{\mathcal C}_{\tau,2} 
\frac{\pi^2}{6}
\left( \frac{1}{M} -M \right)
\, .
\end{align}
%------------------------------------------------------------------------------
The entanglement entropy is $\langle I \rangle={\mathcal C}_{\tau,2} \pi^2/3$. 
The probability distribution of self-information $I$ obtained from the inverse Fourier transform (\ref{iftpi}) is 
%------------------------------------------------------------------------------
\begin{align}
P_\tau(I)
=&
e^{- \langle I \rangle/2}
\delta(I-\langle I \rangle/2)
-i
e^{- I}
\frac{ \theta(I - \langle I \rangle/2) }
{\sqrt{2I/\langle I \rangle-1}}
\nonumber \\
&
\times 
J_1
\left( i \langle I \rangle/2 \sqrt{2I/\langle I \rangle-1} \right)
\, ,
\label{pdfgau}
\end{align}
%------------------------------------------------------------------------------
where $J_n(x)$ is the Bessel function. 
The self-information is almost exponentially distributed and the lower bound is the half of entanglement entropy $I \geq \langle I \rangle/2$. 

\begin{figure}[ht]
\includegraphics[width=0.7 \columnwidth]{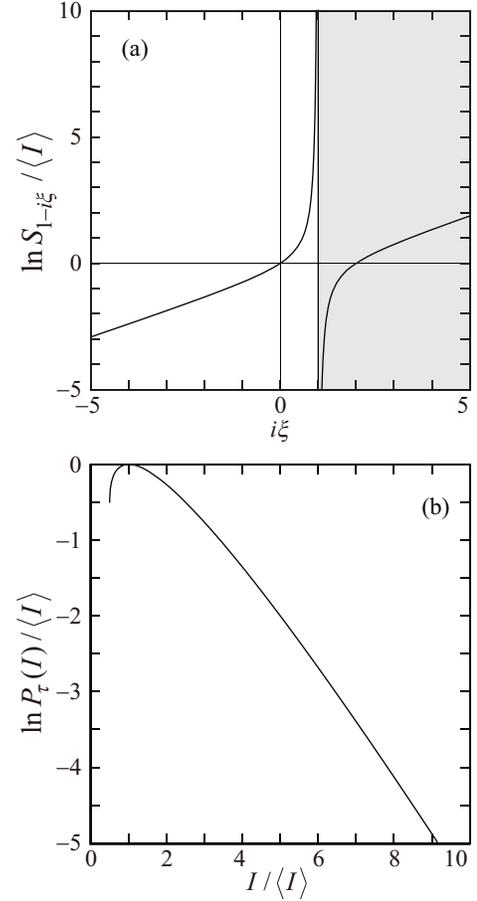}
\caption{
(a) 
The information generating function on the imaginary axis within the Gaussian approximation. 
The information generating function is defined on the domain $- \infty < i \xi < 1$. 
(b) The corresponding probability distributions of self-information for $\langle I \rangle \gg 1$. 
}
\label{pententgau}
\end{figure}

Recall that we are interested in the situation when the source-drain bias voltage is applied and thus the average grows linearly in $\tau$, $\langle I \rangle \propto \tau$. 
In this situation the inverse Fourier transform (\ref{iftpi}) can be done within the saddle-point approximation, i.e., the Legendre-Fenchel transform~\cite{Touchette}, 
%------------------------------------------------------------------------------
\begin{align}
\ln P_\tau(I)
\approx 
\min_{\xi^*} 
\left (
\ln S_{1-i \xi^*} - i \xi^* I
\right )
\, ,
\label{legfentra}
\end{align}
%------------------------------------------------------------------------------
where $\xi^*$ is a pure imaginary number. 
This results in 
%------------------------------------------------------------------------------
\begin{align}
\ln 
P_\tau(I)
\approx
\langle I \rangle
\sqrt{2 I/\langle I \rangle-1}
-I
\, .
\label{pdfgau}
\end{align}
%------------------------------------------------------------------------------
Figures \ref{pententgau} (a) and (b) are the information generating function on the imaginary axis and the probability distribution~(\ref{pdfgau}), respectively. 
The information generating function is defined on the domain $- \infty < i \xi < 1$ [panel (a)]. 
The most probable value, the peak position, is given by the average, i.e., the entanglement entropy [panel (b)]. 
The duality property of Legendre-Fenchel transform relates the slope of the logarithm of information generating function $I^*=\partial_{i \xi} \ln S_{1-i \xi}$ (probability distribution $i \xi^*=-\partial_I \ln P_\tau$) with the argument of the probability distribution $\ln P_\tau(I^*)$ (the information generating function $\ln S_{1-i \xi^*}$)~\cite{Touchette}. 
The logarithm of information generating function is convex and behaves as 
$\ln S_{1 - i \xi} \approx (\langle I \rangle/2) \, i \xi$ for $i \xi \to -\infty$. 
This indicates that the lower bound is $\langle I \rangle/2$. 
It diverges at the boundary $i \xi =1$ [panel (a)]. 
The divergence, in turn, shows that the large fluctuations  are not bounded and $\ln P_\tau (I) \approx -I$ for $I \to \infty$. 
The divergence at $i \xi =1$ implies that the equality (\ref{Jarzynski1}) is not well defined and we have to go beyond the Gaussian approximation, as we will see in the next section. 

%----------------------------------------------------------
%----------------------------------------------------------

\subsection{Limit of long measurement-time}
\label{lml}

In the limit of long measurement time $\tau \to \infty$, the extensive component of ${\mathcal W}_{\sigma \tau}$, i.e., the scaled current cumulant generating function, 
${\mathcal F}_{G \sigma}(\lambda) = \lim_{\tau \to \infty} {\mathcal W}_{\sigma \tau}(\lambda)/\tau$,
is relevant. 
In this limit, Eq.~(\ref{1pcgf}) is calculated as (Appendix \ref{applml}) 
%------------------------------------------------------------------------------
\begin{align}
{\mathcal F}_{G \sigma}(\lambda)
=&
\frac{1}{2 \pi}
\int d \omega
\ln 
\Omega_\lambda(\omega)
\, , 
\label{scgffree}
\\
\Omega_\lambda(\omega)
=&
\frac
{
\tilde{f}_L^+(\omega) 
+ 
\tilde{f}_L^-(\omega) e^{i \lambda}
}
{
{f}_L^+(\omega) 
+ 
{f}_L^-(\omega) e^{i \lambda}
}
\, ,
\end{align}
%------------------------------------------------------------------------------
where a trivial constant is subtracted in order to satisfy the normalization condition ${\mathcal F}_{G \sigma}(0)=0$. 
The effective electron and hole distribution functions, $\tilde{f}_L^+$ and $\tilde{f}_L^-=1-\tilde{f}_L^+$, are bounded to  the interval $[0,1]$. 
The former is the effective transparency from the right lead to the left lead: 
%------------------------------------------------------------------------------
\begin{align}
\tilde{f}_L^+(\omega) 
=
{\mathcal T}(\omega) f_R^+(\omega)
+
{\mathcal R}(\omega) f_L^+(\omega)
\, ,
\end{align}
%------------------------------------------------------------------------------
where the reflection probability is ${\mathcal R}(\omega) = 1-{\mathcal T}(\omega)$. 
At zero temperature, we obtain the scaled current cumulant generating function of the binomial distribution with an energy dependent transmission probability: 
%------------------------------------------------------------------------------
\begin{align}
{\mathcal F}_{G \sigma}(\lambda)
&=
\frac{1}{2 \pi}
\int_{\mu_R}^{\mu_L} d \omega
\ln 
\left[
1+
{\mathcal T}(\omega)
(e^{i \lambda}-1)
\right]
\, . 
\end{align}
%------------------------------------------------------------------------------
Here we consider the positive bias voltage $\mu \equiv \mu_L-\mu_R>0$. 

Using Eq.~(\ref{scgffree}), the spin-resolved R\'enyi entropy is calculated as 
(Appendix \ref{applml}) 
%------------------------------------------------------------------------------
\begin{align}
\ln \frac{S_{\sigma M}}{s_{\sigma M}}
\approx&
\tau 
\sum_{\ell=0}^{M-1}
{\mathcal F}_{G \sigma}(\lambda_\ell)
\nonumber \\
=&
\tau
\int
\frac{d \omega}{2 \pi}
\, 
\ln 
\left(
\frac
{
{\tilde{f}_L^+(\omega)}^M
+ 
{\tilde{f}_L^-(\omega)}^M
}
{
{{f}_L^+(\omega)}^M
+ 
{{f}_L^-(\omega)}^M
}
\right)
\, . 
\label{renent1}
\end{align}
%------------------------------------------------------------------------------
In the limit of zero temperature {\bf (I)} in Sec.~\ref{zerlim}, it becomes
%------------------------------------------------------------------------------
\begin{align}
\ln 
S_{\sigma M}
\approx
\tau
\int_{\mu_R}^{\mu_L} 
\frac{d \omega}{2 \pi}
\, 
\ln 
\left[ 
{\mathcal T}(\omega)^M
+
{\mathcal R}(\omega)^M
\right]
\, . 
\end{align}
%------------------------------------------------------------------------------
In the extended wide-band limit~\cite{HNB}, where the level broadening is large enough, $\mu \ll \Gamma$, or the dot level is far away from the Fermi energy, $|\epsilon_D-\mu_r| \gg \Gamma, \mu$, the transmission probability is energy independent: ${\mathcal T}(\omega) \approx {\mathcal T}(0) = {\mathcal T}$. 
Then the R\'enyi entropy is as follows: 
%------------------------------------------------------------------------------
\begin{align}
S_M
=
\prod_\sigma 
S_{\sigma M}
\approx 
\left(
{\mathcal T}^M
+
{\mathcal R}^M
\right)^{N_{\rm att}}
\, , 
\label{infgenbin}
\end{align}
%------------------------------------------------------------------------------
where $N_{\rm att}=2 \tau \mu/h$ is the number of attempts, i.e., the number of electrons injected into the quantum dot from the left lead during the measurement time $\tau$. 
When $N_{\rm att}$ is a positive integer, Eq.~(\ref{infgenbin}) is the relative information generating function of the binomial distribution~\cite{Guiasu}. 
The entanglement entropy 
%------------------------------------------------------------------------------
\begin{align}
\langle I \rangle
=
-
N_{\rm att}
(
{\mathcal T} \ln {\mathcal T}
+{\mathcal R} \ln {\mathcal R}
)
\, ,
\label{uncentent}
\end{align}
%------------------------------------------------------------------------------
reproduces Ref.~\onlinecite{Beenakker}. 
From Eq.~(\ref{Jarzynski1}), we obtain the number of available states in the subsystem $A$, $\langle e^I \rangle=2^{N_{\rm att}}$. 
The available states are limited to the Fermi window $\mu_R < \omega < \mu_L$, since at zero temperature, the electron states outside this window are empty or occupied. 
The inequality~(\ref{Jarzynski2}) becomes
%------------------------------------------------------------------------------
\begin{align}
N_{\rm att}
(\ln 2
+{\mathcal T} \ln {\mathcal T}
+{\mathcal R} \ln {\mathcal R}
)
\geq 
0
\, . 
\label{Jarzynski2bsc}
\end{align}
%------------------------------------------------------------------------------

\begin{figure}[ht]
\includegraphics[width=0.7 \columnwidth]{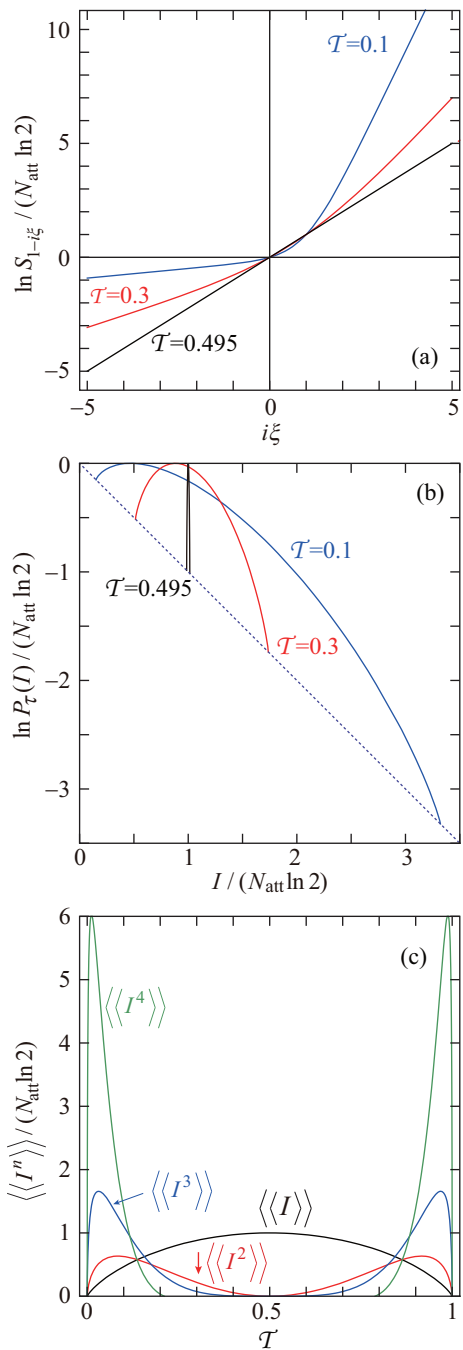}
\caption{
(a) Information generating functions on the imaginary axis and (b) probability distributions for ${\mathcal T}=0.1,0.3$ and $0.495$. 
The upper and lower bounds are on $\ln P_\tau = - I$ (dotted line). 
(c) The lowest 4 cumulants as functions of the transmission probability. 
}
\label{pententbin}
\end{figure}

For the positive integer $N_{\rm att}$, the inverse Fourier transform of Eq.~(\ref{infgenbin}) can be done analytically:
%------------------------------------------------------------------------------
\begin{align}
P_\tau(I)
=&
\sum_{n=0}^{N_{\rm att}}
\frac{N_{\rm att}!}{n! (N_{\rm att}-n)!}
{\mathcal T}^n {\mathcal R}^{N_{\rm att}-n}
\nonumber \\
& \times
\delta (I 
+
n \ln {\mathcal T} + (N_{\rm att}-n) \ln {\mathcal R}
)
\, . 
\label{pdfbinary}
\end{align}
%------------------------------------------------------------------------------
For the binomial process, the distribution is symmetric when we switch the transmission probability and the reflection probability ${\mathcal T} \leftrightarrow {\mathcal R}$. 
Figure \ref{pententbin} (a) shows the information generating function on the imaginary axis for various ${\mathcal T} (<{\mathcal R})$. 
They satisfy the normalization condition $\ln S_1=0$ and the equality ~(\ref{Jarzynski1}) $\ln S_0= N_{\rm att} \ln 2$ independent of the transmission probability. 
Figure \ref{pententbin} (b) shows the corresponding probability distributions calculated within the Legendre-Fenchel transform~(\ref{legfentra})~\cite{Touchette}. 
In the limits of $i \xi \to \infty$ and $i \xi \to -\infty$, the information generating function behaves as 
$\ln S_{1-i \xi} \approx (-N_{\rm att} \ln {\mathcal T}) \, i \xi$ 
and 
$\ln S_{1-i \xi} \approx (-N_{\rm att} \ln {\mathcal R}) \, i \xi$, 
respectively [Fig.~\ref{pententbin} (a)]. 
Therefore, the lower and upper bounds are
$I_{\rm min}=-N_{\rm att} \ln {\mathcal R}$ 
and
$I_{\rm max}=-N_{\rm att} \ln {\mathcal T}$, 
which is consistent with the analytic expression~(\ref{pdfbinary}). 
The upper (lower) bound corresponds to the self-information of a sequence of $N_{\rm att}$ events where all $N_{\rm att}$ injected electrons are transmitted (reflected). 
The probabilities to find $I=I_{\rm min}$ and $I=I_{\rm max}$ are given by 
$P_\tau(I_{\rm min})={\mathcal R}^{N_{\rm att}}=\exp(-I_{\rm min})$ 
and 
$P_\tau(I_{\rm max})={\mathcal T}^{N_{\rm att}}=\exp(-I_{\rm max})$,
respectively. 
Figure \ref{pententbin} (c) shows the lowest 4 cumulants as a function of the transmission probability ${\mathcal T}$. 
The higher cumulants $\langle \! \langle I^n \rangle \! \rangle$ ($n \geq 2$) increase around ${\mathcal T} \approx 0$ or $1$ and vanish at ${\mathcal T}=1/2$. 
The distribution takes a simple form, the delta distribution,  
$P_\tau(I) = \delta (I-N_{\rm att} \ln 2)$, at ${\mathcal T}=1/2$. 

%----------------------------------------------------------
%----------------------------------------------------------

Let us go back to Eq.~(\ref{renent1}) and calculate the average at a finite temperature. 
%------------------------------------------------------------------------------
\begin{align}
\langle I \rangle 
=&
-
\frac{\tau}{2 \pi}
\int d \omega
\sum_\pm 
\tilde{f}_L^\pm(\omega) \ln \tilde{f}_L^\pm(\omega) 
\nonumber \\
&+
S_A
+
\frac{\tau}{2 \pi}
\int d \omega
\sum_\pm 
f_L^\pm(\omega) \ln f_L^\pm(\omega) 
\, .
\label{fintementent}
\end{align}
%------------------------------------------------------------------------------
The first line is the contribution from electrons fluctuating at the boundary between the subsystems $A$ and $B$. 
The second line is the thermodynamic entropy minus the over-counting term.  
Since we can derive $S_{0}=s_{0}$ from Eq.~(\ref{renent1}), we check that the equality (\ref{Jarzynski1}) ensures $\langle e^I \rangle = 2^{N_A}$.

\section{Coulomb interaction}
\label{CI}

\begin{figure}[ht]
\includegraphics[width=0.9 \columnwidth]{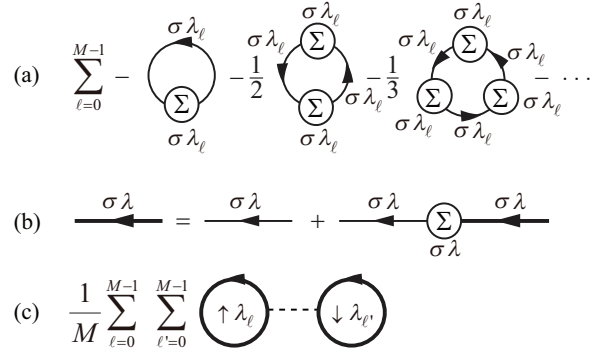}
\caption{
(a) Diagrams for the logarithm of the ``R\'enyi entropy" (\ref{renyi}) of non-interacting electrons  Eq.~(\ref{lincluexp}). 
The thin solid lines represent the bare modified Keldysh Green Function ${\bf g}_{D \sigma}^{\lambda_\ell}$, Eq.~(\ref{mmkgfd}). 
Circles represent the self-energy ${\mathbf \Sigma}_\sigma^{\lambda_\ell}$, Eq.~(\ref{modsel}), which is related to the degrees of freedom of the reservoirs. 
The discretized counting field $\lambda_\ell$ is conserved on a single electron loop. 
(b) Dyson equation for the full modified Keldysh Green function 
${\mathbf G}_{D \sigma}^{\lambda}$, Eq.~(\ref{matdyseqn}). 
(c) Hartree diagram. 
Two loops carry different discretized counting fields, $\lambda_\ell$ and $\lambda_{\ell'}$. \
}
\label{loops}
\end{figure}

Here we calculate the correction induced by the Coulomb interaction by exploiting the Keldysh diagram technique. 
For non-interacting electrons, the series expansion, Eq.~(\ref{lincluexp}), corresponds to the diagrams depicted in Fig.~\ref{loops} (a). 
Each thin solid line represents the bare modified Keldysh Green function ${\mathbf g}_{D \sigma}^{\lambda_\ell}$ and each circle represents the self-energy ${\mathbf \Sigma}_\sigma^{\lambda_\ell}$. 
For non-interacting electrons, the diagrams consist of a single closed electron loop and thus only a single discretized counting field $\lambda_\ell$ appears. 
In the presence of the Coulomb interaction, we have to account for diagrams consisting of more than two electron loops, which carry different discretized counting fields. 
Consequently, the link between the R\'enyi entropy and the full-counting statistics Eq.~(\ref{maires}) does not hold, as we will demonstrate in the following.  

Let us calculate the on-site Coulomb interaction correction up to the lowest order in $U$. 
The first order expansion in $U$ is
%------------------------------------------------------------------------------
\begin{align}
\ln 
\frac{S_M^{(1)}}{s_M}
\approx&
i U
\int_C
dt 
g_{D \uparrow}(t,t)
g_{D \downarrow}(t,t)
\nonumber \\
=&
i U
\sum_{m=1}^M
\sum_{s=\pm}
s
\int_0^\tau \!\! dt 
\, 
g_{D \uparrow}^{ms,ms}(t,t)
\, 
g_{D \downarrow}^{ms,ms}(t,t)
\nonumber \\
=&
\frac{i U}{M}
\sum_{\ell, \ell' =0}^{M-1}
\sum_{s=\pm}
s
\int_0^\tau \!\! dt 
\, 
g_{D \uparrow}^{\lambda_\ell,ss}(t,t)
\, 
g_{D \downarrow}^{\lambda_{\ell'},ss}(t,t)
\, . 
\end{align}
%------------------------------------------------------------------------------
By replacing the bare modified Keldysh Green function ${\mathbf g}_{D \sigma}^\lambda$ with the full modified Keldysh Green function, which follows from the matrix Dyson equation [Fig.~\ref{loops} (b)], 
%------------------------------------------------------------------------------
\begin{align}
{\mathbf G}_{D \sigma}^\lambda
=
{\mathbf g}_{D \sigma}^\lambda
+
{\mathbf g}_{D \sigma}^\lambda
\tau_3 {\mathbf \Sigma}_\sigma^{\lambda} \tau_3
{\mathbf G}_{D \sigma}^\lambda
\, , 
\label{matdyseqn}
\end{align}
%------------------------------------------------------------------------------
(Appendix \ref{fuldotmodkelgrefun}) we obtain the Hartree term, 
%------------------------------------------------------------------------------
\begin{align}
\ln
\frac{S_M^{(1)}}{s_M}
=
\frac{i U}{M}
\sum_{\ell, \ell' =0}^{M-1}
\sum_{s=\pm}
s
\int_0^\tau \!\! dt 
\, 
G_{D \uparrow}^{\lambda_\ell,ss}(t,t)
\, 
G_{D \downarrow}^{\lambda_{\ell'},ss}(t,t)
\, . 
\label{Hartree}
\end{align}
%------------------------------------------------------------------------------
The Hartree diagram is depicted in Fig.~\ref{loops} (c). 
We assign two counting fields $\lambda_\ell$ and $\lambda_{\ell'}$ to two loops and perform summations over both of them. 

%----------------------------------------------------------
%----------------------------------------------------------

The Hartree term in the limit of long measurement time reads 
%------------------------------------------------------------------------------
\begin{align}
\ln 
\frac{S_M^{(1)}}{s_M}
\approx
\tau 
M 
\sum_{\sigma}
U 
\delta n_{\bar{\sigma} M}
n_{\sigma M,q}
\, , 
\end{align}
%------------------------------------------------------------------------------
where $\bar{\sigma}=\uparrow (\downarrow)$ for $\sigma=\downarrow (\uparrow)$. 
The classical component $\delta n_{\bar{\sigma} M}$ and the quantum component $n_{\sigma M,q}$ of dot electron occupancy are given by 
%------------------------------------------------------------------------------
\begin{align}
\delta n_{\bar{\sigma} M}
=&
\sum_{\ell=0}^{M-1}
\frac{
G_{D \bar{\sigma}}^{\lambda_\ell, ++}(t,t)
+
G_{D \bar{\sigma}}^{\lambda_\ell, --}(t,t)
}{2 i M }
\nonumber \\
=&
n_{\bar{\sigma}}
-1/2
+
\delta \tilde{n}_{\bar{\sigma} M}
\, , 
\\
n_{\sigma M,q}
=&
-
\sum_{\ell=0}^{M-1}
\frac{
G_{D \sigma}^{\lambda_\ell, ++}(t,t)
-
G_{D \sigma}^{\lambda_\ell, --}(t,t)
}{M}
\nonumber \\
=&
\frac{1}{M \tau}
\partial_{\epsilon_D}
\ln \frac{S_{\sigma M}}{s_{\sigma M}}
\, . 
\end{align}
%------------------------------------------------------------------------------
The Hartree term is interpreted as a correction caused by $M$-dependent renormalization of the dot level, 
$\epsilon_D \to \epsilon_D+U \delta n_{\bar{\sigma} M}$. 
For $M=1$, the quantum component vanishes $n_{\sigma 1,q}=0$, as we see from Eq.~(\ref{renent1}). 
The classical component becomes 
$\delta n_{\bar{\sigma} 1}=n_{\bar{\sigma}} -1/2$. 
The spin-resolved occupancy is as follows [hereafter, we concentrate on the zero temperature limit {\bf (I)} in Sec.~\ref{zerlim}]: 
%------------------------------------------------------------------------------
\begin{align}
n_{\bar{\sigma}}
=
\frac{1}{2}
+
\sum_{r=L,R}
\frac{\Gamma_r}{\pi \Gamma}
\tan^{-1}
\left(
\frac{\mu_r - \epsilon_D}{\Gamma/2}
\right)
\, .
\end{align}
%------------------------------------------------------------------------------
The $M$-dependent correction is 
%------------------------------------------------------------------------------
\begin{align}
\delta \tilde{n}_{\bar{\sigma} M}
=
\frac{\Gamma_L - \Gamma_R}{2 \Gamma}
\int_{\mu_R}^{\mu_L} 
d \omega
\rho(\omega)
\left(
\frac{
{\mathcal R}(\omega)^{M-1}
}{
{\mathcal T}(\omega)^{M}
+
{\mathcal R}(\omega)^{M}
}
-1
\right)
\, . 
\end{align}
%------------------------------------------------------------------------------

It turns out that the Hartree term affects rare events, which correspond to $i \xi \to \pm \infty$ limits after the analytic continuation $M \to 1 - i \xi$. 
When the condition ${\mathcal T}(\omega) > {\mathcal R}(\omega)$ is satisfied in the Fermi window $\mu_R < \omega <\mu_L$, the classical component in each limit reads
%------------------------------------------------------------------------------
\begin{align}
\lim_{i \xi \to - \infty }
\delta n_{\bar{\sigma}, 1-i \xi}
=&
\sum_{r=L,R}
\frac{1}{2 \pi}
\tan^{-1}
\frac{\mu_r - \epsilon_D}{\Gamma/2}
=
\delta n_{\bar{\sigma}}^{{\rm sym}} \, ,
\\
\lim_{i \xi \to \infty }
\delta n_{\bar{\sigma}, 1-i \xi}
=&
\sum_{r=L,R}
\frac{1}{2 \pi}
\tan^{-1}
\frac{\mu_r - \epsilon_D}{(\Gamma_r - \Gamma_{\bar{r}})/2}
+
\delta n_{\bar{\sigma}}^{{\rm sym}}
\, , 
\end{align}
%------------------------------------------------------------------------------
where $\bar{r}=L (R)$ for $r=R (L)$. 
The former is the dot occupancy (subtracted by $1/2$) for the symmetric coupling $\Gamma_L = \Gamma_R$. 
The classical component $\delta n_{\bar{\sigma}, 1- i \xi}$ can be simplified further in the case that the dot level is between two chemical potentials, $\mu_R < \epsilon_D < \mu_L$, near the symmetric coupling, $|\Gamma_L - \Gamma_R|/2 \ll |\mu_r - \epsilon_D| \ll \Gamma/2 $: 
%------------------------------------------------------------------------------
\begin{align}
\delta n_{\bar{\sigma}, 1- i \xi}
\approx&
\left \{
\begin{array}{cc}
0 & (i \xi \to - \infty) \\
0 & (i \xi = 0) \\
{\rm sgn} (\Gamma_L - \Gamma_R)/2 & (i \xi \to \infty)
\end{array}
\right. 
\, .
\end{align}
%------------------------------------------------------------------------------
This indicates that for the upper bound $i \xi \to \infty$, where all injected electrons are reflected, each electron observes that the dot is fully occupied for $\Gamma_L > \Gamma_R$ or empty for $\Gamma_R > \Gamma_L$. 
In contrast, for the lower bound $i \xi \to - \infty$, where all electrons are transmitted, each electron observes that the dot is half occupied. 

%----------------------------------------------------------
\begin{figure}[ht]
\includegraphics[width=0.7 \columnwidth]{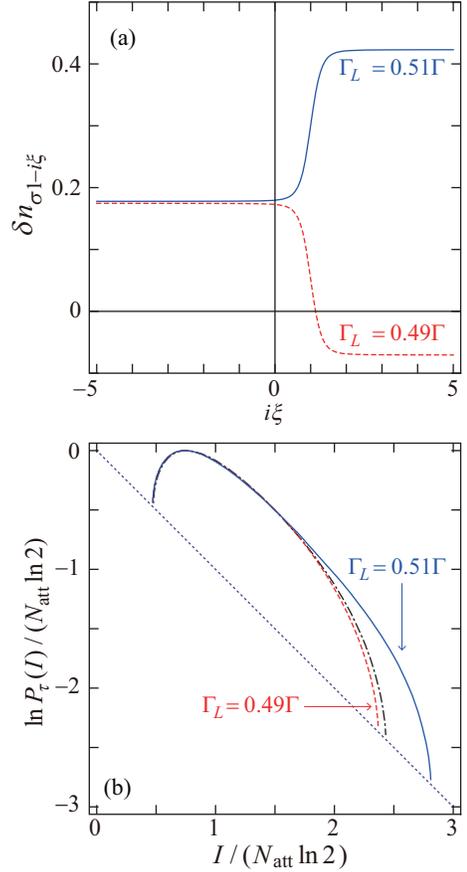}
\caption{
(a) Dot electron occupancies as functions of $i \xi$ and (b) the corresponding probability distributions for $\Gamma_L=0.51 \Gamma$ (solid line) and $\Gamma_L=0.49 \Gamma$ (dashed line). 
The dot level is off the symmetric point $\mu_L-\epsilon_D \neq \epsilon_D - \mu_R$. 
The limit $i \xi \to \infty$ ($i \xi \to -\infty$) corresponds to a rare event in which all electrons are reflected (transmitted). 
The dot-dashed line in panel (b) is the result without Coulomb interaction $U=0$. 
Parameters: 
$\mu_L = - \mu_R=-\epsilon_D=0.5 \Gamma$, $U = 0.1 \Gamma$. 
}
\label{pententHartree}
\end{figure}
%----------------------------------------------------------

Figure~\ref{pententHartree} (a) shows the $\xi$ dependence of the dot occupancy $\delta n_{\sigma \, 1 - i \xi}$ for the nearly symmetric coupling  $\Gamma_L \approx \Gamma_R$. 
At $i \xi \to -\infty$, where all electrons are transmitted, the dot occupancy is slightly modified as compared with that at $i \xi=0$. 
On the other hand, at $i \xi \to \infty$, where all electrons are reflected, the dot occupancy is enhanced ($\Gamma_L > \Gamma_R$) or suppressed ($\Gamma_L < \Gamma_R$). 
Figure~\ref{pententHartree} (b) shows the corresponding probability distributions. 
The upper bound is modified as compared with that without Coulomb interaction (dot-dashed line) although the peak position, i.e., the entanglement entropy, is almost unchanged.

\section{Summary}
\label{Sum}

In summary, we studied statistical properties of information content in the presence of a Coulomb interaction. 
We calculated the R\'enyi entropy of a \YU{positive} integer order $M$ by exploiting the multi-contour Keldysh Green function. 
We demonstrated that at zero temperature, the discrete Fourier transform of the multi-contour Keldysh Green function is compatible with the modified Keldysh Green function introduced previously in the context of full-counting statistics. 
For non-interacting electrons, we relate the current cumulant generating function of the full-counting statistics, the R\'enyi entropy and the entanglement entropy without relying on the correlation matrix. 
We further calculate the probability distribution of self-information by the inverse Fourier transform of the information generating function obtained by the analytic continuation $M \to 1-i \xi$. 
Within the Hartree approximation, we demonstrate that, in the vicinity of the perfect transmission, the dot occupancy is modified for rare events. 
Consequently, the upper bound of the probability distribution of self-information is modified. 
We point out the equality reminiscent to the Jarzynski equality, from which the upper bound of the entanglement entropy could be obtained. 

\YU{
In short, we feel there are two important and concrete messages in the present paper; 
(1) The discrete Fourier transform of the multi-contour Keldysh Green function in the replicated Keldysh space provides a feasible way to calculate the R\'enyi entropy starting from a microscopic model Hamiltonian. 
It enables us to deal with electron interactions by a slight extension of traditional diagrammatic techniques. 
(2) The R\'enyi entropy may contain the information on the fluctuations of information content beyond the average value, the entanglement entropy. 
The concept of the probability distribution of self-information provides one way to interpret the meaning of the R\'enyi entropy. 
}
It could be interesting to consider the conditional joint probability distribution of energy~~\cite{WSU} and information content and analyze the possible connection between these quantities~\cite{Ansari2}.

We thank Dmitry Golubev, Ryuichi Shindou and Kazutaka Takahashi for their valuable input. 
This work was supported by JSPS KAKENHI grants (grants no. 26400390 and no. 26220711).

\begin{appendix}

\begin{widetext}

\section{Bloch-De Dominicis theorem for the normalized expectation value}
\label{KGF}

We demonstrate the Bloch-De Dominicis theorem~\cite{KTH} for the normalized expectation value (\ref{nev}) through calculations of the following 4-th order term as an example (in this section, we neglect the subscripts $k$ and $\sigma$ for simplicity). 
%------------------------------------------------------------------------------
\begin{align}
(-i)^4
\left \langle
a_{L}(t_4)_I^\dagger
d(t_4)_I 
d(t_3)_I^\dagger 
a_{L}(t_3)_I
d(t_2)_I^\dagger
a_{R}(t_2)_I
a_{R}(t_1)_I^\dagger
d(t_1)_I
\right \rangle_M
\, , 
\label{4a}
\end{align}
%------------------------------------------------------------------------------
where $t_1 > t_2$ ($t_1, t_2 \in C_{1,-}$), $t_3 \in C_{m_0,+}$, and $t_4 \in C_{m_0+m_1+1,-}$ [Fig. \ref{bdd}]. 
This is calculated as 
%------------------------------------------------------------------------------
\begin{align}
{\rm Tr}
\left[
(\rho_{L}\rho_{D})^{M-m_1-m_0}
a_{L}^\dagger
d
(\rho_{L}\rho_{D})^{m_1}
d^\dagger 
a_{L}
(\rho_{L}\rho_{D})^{m_0}
\rho_{R \,1}
d^\dagger
a_{R \, 1}
a_{R \, 1}^\dagger
d
\right]
e^{
-i \epsilon_D (t_4-t_3-t_2+t_1)
-i \epsilon_{L} (t_3-t_4)
-i \epsilon_{R} (t_2-t_1)
}
/s_{LM} s_{DM}
\, , 
\label{bdd1}
\end{align}
%------------------------------------------------------------------------------
The density matrices of the replicated subsystem $B$, $\rho_{R \, m}$ disappear except for $\rho_{R \, 1}$. 
Following the standard procedure~\cite{KTH}, we calculate the trace in Eq.~(\ref{bdd1}) as
%------------------------------------------------------------------------------
\begin{align}
&
{\rm Tr}
\left[
{\rho_D}^{M-m_1-m_0}
d
{\rho_D}^{m_1}
d^\dagger 
{\rho_D}^{m_0}
\rho_{R \,1}
d^\dagger
a_{R \, 1}
a_{R \, 1}^\dagger
d
\right]
f_{L,M-m_1}(\epsilon_{L})
s_{LM}
\nonumber \\
=&
{\rm Tr}
\left[
{\rho_D}^{M-m_1-m_0}
d
{\rho_D}^{m_1}
d^\dagger 
{\rho_D}^{m_0}
d^\dagger
d
\right]
f_{R}^-(\epsilon_{R})
f_{L,M-m_1}(\epsilon_{L})
s_{LM}
\nonumber \\
=&
\left[
f_{D,m_1}(\epsilon_D)
f_{D,M}(\epsilon_D)
-
f_{D,M-m_0}(\epsilon_D)
f_{D,m_0+m_1}(\epsilon_D)
\right]
f_{R}^-(\epsilon_{R})
f_{L,M-m_1}(\epsilon_{L})
s_{LM}
s_{DM}
\, , 
\end{align}
%------------------------------------------------------------------------------
where the modified Fermi distribution function is defined in Eq.~(\ref{mfd}). 
Then Eq.~(\ref{4a}) is expressed with multi-contour Keldysh Green functions~(\ref{repkelgf}) as 
%------------------------------------------------------------------------------
\begin{align}
&g_D^{1 \, -,1 \, -}(t_1,t_2)
g_{R}^{1 \, -,1 \, -}(t_2,t_1)
g_D^{m_0+m_1+1 \, -,m_0 \, +}(t_4,t_3)
g_{L}^{m_0 \, +,m_0+m_1+1 \, -}(t_3,t_4)
\nonumber 
\\
&-
g_D^{1 \, -,m_0 \, +}(t_1,t_3)
g_{L}^{m_0 \, +,m_0+m_1+1 \, -}(t_3,t_4)
g_D^{m_0+m_1+1 \, -,1 \, -}(t_4,t_2)
g_{R}^{1 \, - ,1 \, -}(t_2,t_1)
\, , 
\label{4aa}
\end{align}
%------------------------------------------------------------------------------
which proves the Bloch-De Dominicis theorem. 
The first and second terms correspond to the diagrams depicted in Fig.~\ref{bdd} (a) and (b), respectively. 

%----------------------------------------------------------
\begin{figure}[ht]
\includegraphics[width=0.5 \columnwidth]{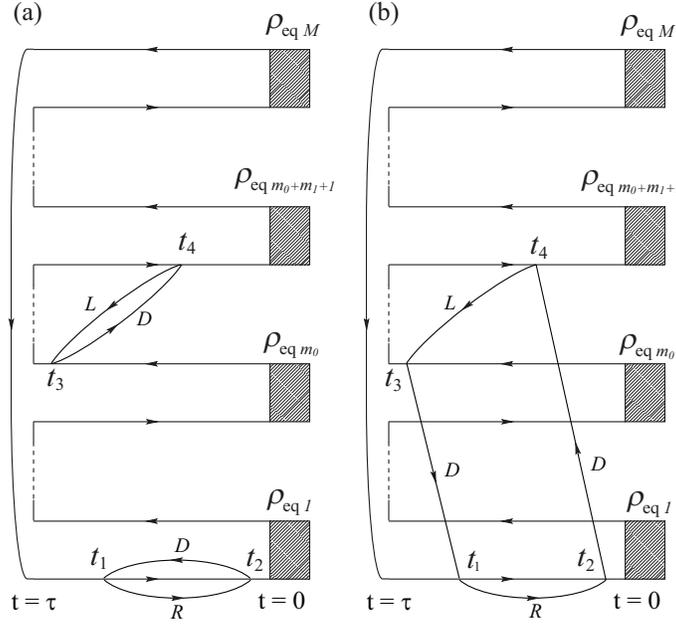}
\caption{
Diagrams of the 4-th order term, Eq.~(\ref{4a}). 
}
\label{bdd}
\end{figure}
%----------------------------------------------------------

\section{Discrete Fourier transform}
\label{sec:dft}

In this section, we summarize  the multi-contour Keldysh Green function introduced in Ref.~\onlinecite{Nazarov}. 
Let us calculate Eq.~(\ref{gf}) for $s=+$, $s=-$ and $m > m'$ as an example. 
Paying attention that the replicated equilibrium density matrices $\rho_{{\rm eq} \, m}$ ($m=1,\cdots, M$) also obey the contour ordering operator $T_C$, Eq.~(\ref{gf}) is calculated as 
%------------------------------------------------------------------------------
\begin{align}
g_{L k \sigma}^{m+,m'-}(t,t')
=&
-i 
{\rm Tr}
\left[
T_C
a_{L k \sigma}(t_{ms})_I
a_{L k \sigma}^\dagger(t_{m's'}')_I
\rho_{{\rm eq} \, M}
\cdots
\rho_{{\rm eq} \, 1}
\right]/s_M
\nonumber \\
=&
-i 
\, 
{\rm Tr}
\left[
{\rho_{L}}^{M-m+m'-1}
a_{L k \sigma}(t)_I
{\rho_{L}}^{m'-m+1}
a_{L k \sigma}^\dagger(t')_I
\right]
/s_{L M}
=
-i f_{L,m-m'+1}(\epsilon_{Lk}) e^{-i \epsilon_{Lk} (t-t')}
\, , 
\end{align}
%------------------------------------------------------------------------------
where
$
s_{L M}=\prod_\sigma s_{L \sigma M}
$. 
Here, the Fermi distribution function is modified and $M$-dependent;
%------------------------------------------------------------------------------
\begin{align}
f_{r,m}(\epsilon)
=
\frac{e^{-m \beta (\epsilon-\mu_r)}}{1+e^{-M \beta (\epsilon-\mu_r)}}
\, . 
\label{mfd}
\end{align}
%------------------------------------------------------------------------------
We calculate the other components in the same way and obtain a $2 \times 2$ sub-matrix of a $2M \times 2M$ multi-contour Keldysh Green function matrix connecting branches $C_{m, \, \pm}$ and $C_{m' \, \pm}$ as 
%------------------------------------------------------------------------------
\begin{align}
\left[
{\mathbf g}_{L k \sigma}(t,t')
\right]_{m,m'}
=&
\left[
\begin{array}{cc}
g_{L k \sigma}^{m+,m'+}(t,t') & g_{L k \sigma}^{m+,m'-}(t,t') \\
g_{L k \sigma}^{m-,m'+}(t,t') & g_{L k \sigma}^{m-,m'-}(t,t')
\end{array}
\right]
=
-i 
e^{-i \epsilon_{Lk} (t-t')}
\nonumber \\
&\times
\left \{
\begin{array}{cc}
\left[
\begin{array}{cc}
f_{L,m-m'}(\epsilon_{Lk}) & f_{L,m-m'+1}(\epsilon_{Lk}) \\
f_{L,m-m'-1}(\epsilon_{Lk}) & f_{L,m-m'}(\epsilon_{Lk})
\end{array}
\right]
& 
(m > m')
\\
\left[
\begin{array}{cc}
f_{L,0}(\epsilon_{Lk}) \theta(t-t') - f_{L,M}(\epsilon_{Lk}) \theta(t'-t) 
& 
f_{L,1}(\epsilon_{Lk})
\\
-
f_{L,M-1}(\epsilon_{Lk})
& f_{L,0}(\epsilon_{Lk}) \theta(t'-t) - f_{L,M}(\epsilon_{Lk}) \theta(t-t')
\end{array}
\right]
&
(m'=m)
\\
\left[
\begin{array}{cc}
-f_{L,M+m-m'}(\epsilon_{Lk}) & -f_{L,M+m-m'+1}(\epsilon_{Lk}) \\
-f_{L,M+m-m'-1}(\epsilon_{Lk}) & -f_{L,M+m-m'}(\epsilon_{Lk})
\end{array}
\right]
&
(m<m')
\end{array}
\right.
. 
\label{repkelgf}
\end{align}
\end{widetext}
%------------------------------------------------------------------------------
The multi-contour Keldysh Green function (\ref{repkelgf}) possesses the discrete translational symmetry:
%------------------------------------------------------------------------------
\begin{align}
\left[
{\mathbf g}_{L k \sigma}
\right]_{m-m'}
\equiv &
\left[
{\mathbf g}_{L k \sigma}
\right]_{m,m'}
\, , 
\label{distrasyn}
\end{align}
%------------------------------------------------------------------------------
where $m-m'=-M+1, \cdots, M-1$. 
It also satisfies the anti-periodic boundary condition: 
%------------------------------------------------------------------------------
\begin{align}
\left[
{\mathbf g}_{L k \sigma}
\right]_{n}
=&
-
\left[
{\mathbf g}_{L k \sigma}
\right]_{n+M}
\, , 
\label{antperboucon}
\end{align}
%------------------------------------------------------------------------------
where $n=-M+1, \cdots, -1$. 
Therefore, it is expedient to set 
$\left[{ \mathbf g}_{L k \sigma} \right]_{\pm M} = - \left[ {\mathbf g}_{L k \sigma} \right]_{0}$
and continue the domain of the function $\left[ {\mathbf g}_{L k \sigma} \right]_{n}$ periodically  with the period $2M$~\cite{AGD}. 
The discrete Fourier transform and the inverse discrete Fourier transform read 
%------------------------------------------------------------------------------
\begin{align}
{\mathbf g}_{L k \sigma}[j]
=&
\frac{1}{2}
\sum_{n=0}^{2 M-1}
\left[
{\mathbf g}_{L k \sigma}
\right]_{n}
e^{i 2\pi j n/(2M)}
\, , 
\label{dft2}
\\
\left[
{\mathbf g}_{L k \sigma}
\right]_{n}
=&
\frac{1}{M}
\sum_{j=0}^{2 M-1}
{\mathbf g}_{L k \sigma}[j]
e^{-i 2\pi j n/(2M)}
\, . 
\label{idft2}
\end{align}
%------------------------------------------------------------------------------
By using the anti-periodic boundary condition (\ref{antperboucon}), we rewrite Eq.~(\ref{dft2}) and demonstrate that an even $j$ component  vanishes:
%------------------------------------------------------------------------------
\begin{align}
{\mathbf g}_{L k \sigma}[j]
=
\sum_{n=0}^{M-1}
\frac{1+(-1)^j}{2}
\left[
{\mathbf g}_{L k \sigma}
\right]_{n}
e^{i \pi j n/M}
\, , 
\end{align}
%------------------------------------------------------------------------------
Then, by setting ${\mathbf g}_{L k \sigma}^{\lambda_\ell} = {\mathbf g}_{L k \sigma}[2 \ell+1]$ ($\ell=0, \cdots ,M-1$), Eqs.~(\ref{dft2}) and (\ref{idft2}) are reduced to Eqs.~(\ref{dft}) and (\ref{idft}). 

In order to calculate the discrete Fourier transform (\ref{dft}), we fix $m'$ and rewrite the summation as 
%------------------------------------------------------------------------------
\begin{align}
{\mathbf g}_{L k \sigma}^{\lambda_\ell}(t,t')
=&
\sum_{m=1}^{M}
\left[
{\mathbf g}_{L k \sigma}(t,t')
\right]_{m,m'}
x_\ell^{m-m'}
\, , 
\end{align}
%------------------------------------------------------------------------------
where $x_\ell = e^{i \pi (2 \ell+1)/M}$. 
The discrete Fourier transform of $g_{L k \sigma}^{m+,m'-}$ is calculated as 
%------------------------------------------------------------------------------
\begin{widetext}
\begin{align}
g_{L k \sigma}^{\lambda_\ell,+-}(t,t')
=&
-i e^{-i \epsilon_{k}(t-t')}
(
-
f_{L,M-m'-2}
x_\ell^{1-m'}
-
\cdots 
-
f_{L,M}
x_\ell^{-1}
+
f_{L,1}
+
f_{L,2-m'}
x_\ell^{1}
+
\cdots
+
f_{L,M-m'+1}
x_\ell^{M-m'}
)
\nonumber \\
=&
-i e^{-i \epsilon_{k}(t-t')}
\sum_{j=0}^{M-1}
f_{L,j+1}
x_\ell^j
=
-i e^{-i \epsilon_{k}(t-t')}
f_{L,\lambda_\ell}^+(\epsilon_{Lk})
e^{i \lambda_\ell}
\, . 
\end{align}
%------------------------------------------------------------------------------
The causal component is calculated as 
%------------------------------------------------------------------------------
\begin{align}
g_{L k \sigma}^{\lambda_\ell,++}(t,t')
=&
-i e^{-i \epsilon_{k}(t-t')}
(
-f_{L,M-m'+1}
x_\ell^{1-m'}
-
\cdots 
-
f_{L,M-1}
x_\ell^{-1}
+
f_{L,0} \theta(t-t')
-
f_{L,M} \theta(t'-t)
+
f_{L,1}
x_\ell^{1}
+
\cdots
\nonumber \\
&+
f_{L,M-m'}
x_\ell^{M-m'}
)
=
-i e^{-i \epsilon_{k}(t-t')}
\left[
\theta(t-t')
\sum_{j=0}^{M-1}
f_{L,j} x_\ell^{j}
+
\theta(t'-t)
\sum_{j=0}^{M-1}
f_{L,j+1} x_\ell^{j+1}
\right]
\nonumber \\
=&
-i e^{-i \epsilon_{k}(t-t')}
\left[
f_{L,\lambda_\ell}^-(\epsilon_{k})
\theta(t-t')
-
f_{L,\lambda_\ell}^+(\epsilon_{k})
\theta(t'-t)
\right]
\, . 
\end{align}
%------------------------------------------------------------------------------
For the other two components of Eq.~(\ref{mkgfl}), we repeat the same calculations  and obtain Eq.~(\ref{mmkgf}).

\section{Full modified Keldysh Green function for the quantum dot}
\label{fuldotmodkelgrefun}

Here, we solve the matrix Dyson equation~(\ref{matdyseqn}) in the limit of $\tau \to \infty$. 
The Fourier transform of the bare modified Keldysh Green function of the left lead~(\ref{mmkgf}) is 
%------------------------------------------------------------------------------
\begin{align}
{\mathbf g}_{L k \sigma}^\lambda(\omega)
=&
\int d (t-t') e^{i \omega (t-t')}
{\mathbf g}_{L k \sigma}^\lambda(t,t')
=
\left[
\begin{array}{cc}
\frac{ f_{L,\lambda}^-(\epsilon_{Lk}) }
{\omega + i \eta -\epsilon_{Lk}}
+
\frac{ f_{L,\lambda}^+(\epsilon_{Lk}) }
{\omega - i \eta -\epsilon_{Lk}}
&
-2 \pi i
f_{L,\lambda}^+(\omega)
e^{i \lambda}
\delta(\omega - \epsilon_{Lk})
\\
2 \pi i 
f_{L,\lambda}^-(\omega)
e^{-i \lambda}
\delta(\omega - \epsilon_{Lk})
&
-
\frac{ f_{L,\lambda}^-(\epsilon_{Lk}) }
{\omega - i \eta -\epsilon_{Lk}}
-
\frac{ f_{L,\lambda}^+(\epsilon_{Lk}) }
{\omega + i \eta -\epsilon_{Lk}}
\end{array}
\right]
\, ,
\label{mmkgfome}
\end{align}
%------------------------------------------------------------------------------
where $\eta$ is a positive infinitesimal and the delta function is defined as 
$\delta(\omega)=\eta/[\pi (\omega^2 + \eta^2)]$. 
The modified self-energy~(\ref{modsel}) is~\cite{UGS,UEUA,GK,U,SU,US,Esposito,HNB}, 
%------------------------------------------------------------------------------
\begin{align}
\Sigma_{\sigma}^\lambda(\omega)
=
-i 
\sum_{r=L,R}
\frac{\Gamma_r}{2}
\left[
\begin{array}{cc}
1-2 f_{r, \lambda_r}^+(\omega) & 2 f_{r, \lambda_r}^+(\omega) e^{i \lambda_r} \\
- 2 f_{r, \lambda_r}^-(\omega) e^{-i \lambda_r} & 
1-2 f_{r, \lambda_r}^+(\omega)
\end{array}
\right]
\, ,
\;\;\;\;
(\lambda_L=\lambda, \lambda_R=0)
\, .
\end{align}
%------------------------------------------------------------------------------
The bare modified Keldysh Green function of the quantum dot is given in a similar form as Eq.~(\ref{mmkgf}):
%------------------------------------------------------------------------------
\begin{align}
{\mathbf g}_{D \sigma}^\lambda(t,t')
=
-i e^{-i \epsilon_{D}(t-t')}
\left[
\begin{array}{cc}
f_{D,\lambda}^-(\epsilon_{D})
\theta (t-t')
-
f_{D,\lambda}^+(\epsilon_{D})
\theta (t'-t)
&
f_{D,\lambda}^+(\epsilon_{D}) e^{i \lambda} \\
-
f_{D,\lambda}^-(\epsilon_{D}) e^{-i \lambda}
&
f_{D,\lambda}^-(\epsilon_{D})
\theta (t'-t)
-
f_{D,\lambda}^+(\epsilon_{D})
\theta (t-t')
\end{array}
\right]
\, . 
\label{mmkgfd}
\end{align}
%------------------------------------------------------------------------------
The full modified Keldysh Green function obtained after solving the matrix Dyson equation~(\ref{matdyseqn}) is
%------------------------------------------------------------------------------
\begin{align}
{\mathbf G}_{D \sigma}^\lambda(\omega)
&=
\frac{\Omega_\lambda(\omega)^{-1}}{(\omega-\epsilon_D)^2 + \Gamma^2/4}
\left[
\begin{array}{cc}
\omega-\epsilon_D - i \sum_r \Gamma_r [1/2 - f_{r,\lambda_r}^+(\omega)] & 
-i \sum_r \Gamma_r f_{r,\lambda_r}^+(\omega) e^{i \lambda_r} \\
 i \sum_r \Gamma_r f_{r,\lambda_r}^-(\omega) e^{-i \lambda_r} & 
\epsilon_D-\omega - i \sum_r \Gamma_r [1/2 - f_{r,\lambda_r}^+(\omega)]
\end{array}
\right]
\, , 
\label{hgf}
\\
\Omega_\lambda(\omega)
=&
1
+
{\mathcal T}(\omega)
[
f_{L,\lambda}^+ f_{R}^-
(e^{i \lambda}-1)
+
f_{R}^+
f_{L,\lambda}^- 
(e^{-i \lambda}-1)
]
=
\frac
{
\tilde{f}_L^+(\omega) 
+ 
\tilde{f}_L^-(\omega) e^{i \lambda}
}
{
{f}_L^+(\omega) 
+ 
{f}_L^-(\omega) e^{i \lambda}
}
\, . 
\label{omega}
\end{align}
%------------------------------------------------------------------------------
\end{widetext}

\section{Details of calculations in Sec.~\ref{lml}}
\label{applml}

By using the full modified Keldysh Green function, Eq.~(\ref{matdyseqn}), Eq.~(\ref{1pcgf}) can be rewritten as 
%------------------------------------------------------------------------------
%\begin{align}
${\mathcal W}_{\sigma \tau}(\lambda)
=
{\rm Tr}
\ln 
{{\mathbf G}_{D \sigma}^\lambda}^{-1}
{{\mathbf g}_{D \sigma}^\lambda}
$. 
%\end{align}
%------------------------------------------------------------------------------
By using Eqs.~ (\ref{mmkgfome}) and (\ref{hgf}), the limit of $\tau \to \infty$ is calculated as  
%------------------------------------------------------------------------------
\begin{align}
{\mathcal W}_{\sigma \tau}(\lambda)
\approx&
\frac{\tau}{2 \pi}
\int d \omega
\ln 
\frac{
{\det {\mathbf G}_{D \sigma}^\lambda(\omega)}^{-1} }{
{\det {\mathbf G}_{D \sigma}^{\lambda=0}(\omega)}^{-1}
}
\nonumber \\
&+
\frac{\tau}{2 \pi}
\int d \omega
\ln 
\frac{
{\det {\mathbf G}_{D \sigma}^{\lambda=0}(\omega)}^{-1} }{
{\det {\mathbf g}_{D \sigma}^{\lambda}(\omega)}^{-1}
}
\nonumber \\
=&
\frac{\tau}{2 \pi}
\int d \omega
\ln 
\Omega_\lambda(\omega)
+
\tau \left( \frac{\Gamma}{2}+\eta \right)
\, , 
\end{align}
%------------------------------------------------------------------------------
which gives Eq.~(\ref{scgffree}) except for a constant $\tau \Gamma/2$. 

The summation over the discretized counting field, Eq.~(\ref{renent1}), can be done for $\tilde{f}_L^-(\omega) < \tilde{f}_L^+(\omega)$ as follows: 
%------------------------------------------------------------------------------
\begin{align}
&
\sum_{\ell=0}^{M-1}
\ln 
\left(
\tilde{f}_L^+(\omega) +
\tilde{f}_L^-(\omega) e^{i \lambda_\ell}
\right)
\nonumber \\
=&
M 
\ln {\tilde{f}_L^+}
-
\sum_{n=1}^\infty
\frac{1}{n}
\left( \frac{-\tilde{f}_L^-}{\tilde{f}_L^+} \right)^n
\sum_{\ell=0}^{M-1}
e^{i \lambda_\ell n}
\nonumber \\
=&
M \ln {\tilde{f}_L^+}
-
\sum_{n=1}^\infty
\frac{1}{n}
\left( \frac{-\tilde{f}_L^-}{\tilde{f}_L^+} \right)^n
M (-1)^{n (1-1/M)}
\delta_{n ,k M}
\nonumber
\\
=&
\ln 
\left(
{\tilde{f}_L^+(\omega)}^M
+
{\tilde{f}_L^-(\omega)}^M
\right)
\, , 
\end{align}
%------------------------------------------------------------------------------
where $k$ is an integer. 
For $\tilde{f}_L^-(\omega) > \tilde{f}_L^+(\omega)$, one can repeat similar calculations and obtain the same result.

\end{appendix}


\begin{thebibliography}{99}
\bibitem{note1}
\YU{
The R\'enyi entropy of order $M$ is introduced to be extensive by using logarithm as $(\ln {\rm Tr}_A
\left[
{\rho_A}^M
\right]
)/(1-M)$ for $M \geq 0$ and $M \neq 1$ , see, e.g. Ref.~\onlinecite{Cover}. 
In the present paper, we obey the terminology in Ref.~\onlinecite{Nazarov}. 
}
\bibitem{noteres1}
\YU{
Let us explain a trivial example, non-interacting electrons decoupled from the subsystem B. 
At zero temperature, we have a pure state density matrix 
$\rho_A = |{\rm FS} \rangle \langle {\rm FS}|$, 
where $|{\rm FS} \rangle$ is a unique ground-state many-body wave function; 
At the ground state, electrons fill up to the Fermi energy. 
In this case the rank of ${\rho_A}^0$ is 1. 
If the temperature is finite, ${\rho_A}^0={\mathbf 1}$, which is the unit matrix defined in the whole many-body Fock space of the subsystem $A$. 
Then the rank of ${\rho_A}^0$ is $2^{N_A}$, where $N_A$ is the number of electron orbits in the subsystem $A$. 
}
\end{thebibliography}
\end{document}